\documentclass[aps,preprint,groupedaddress]{revtex4}
\usepackage{graphicx}
\usepackage{dcolumn}
\usepackage{amsmath}

\def\be#1{\begin{equation}\label{#1}}
\def\ee{\end{equation}}
\def\bea#1{\begin{eqnarray}\label{#1}}
\def\eea{\end{eqnarray}}
\def\Bea{\begin{eqnarray*}}
\def\Eea{\end{eqnarray*}}
\def\sp{\hspace{.5em}}
\def\sph{\hspace{.25em}}
\def\spp{\sp\sp}
\def\Eq#1{Eq.(\ref{#1})}
\def\Fig#1{Fig.(\ref{#1})}

\def\no{\nonumber \\}
\def\tbf#1{\textbf{#1}}
\def\trm#1{\textrm{#1}}
\def\tit#1{\textit{#1}}

\def\displ{\displaystyle}
\def\m#1{\mathbf{#1}}

\def\mbf#1{\mbox{{\boldmath $#1$}}}
\def\half{\frac{1}{2}}
\def\ket#1{|#1\rangle}

\def\expm2piOmega{e^{-2\pi\Omega}}

\def\sp{\hspace{.25em}}
\def\d#1#2{d^{\sp(#1)}_{\Omega,#2\vec{k}_\perp}}

\def\b#1#2{b^{\sp(#1)}_{\Omega,#2\vec{k}_\perp}}

\def\a#1{a_{#1\vec{k}_\perp,k^3}}

\def\bfe#1{\mathbf{e}_{\hat{#1}}}

\def\e#1#2{e_{\hat{#1}}^{\spp #2}}
\def\inve#1#2{e_{#1}^{\spp \hat{#2}}}
\def\inveT#1#2{e^{\hat{#1}}_{\spp #2}}

\def\spinc#1#2#3{\omega_{#1 \spp #3}^{\spp #2}}
\def\spincup#1#2#3{\omega_{#1}^{\spp #2\,#3}}
\def\spincdn#1#2#3{\omega_{#1\,#2\,#3}}

\def\ha{\hat{a}}
\def\hb{\hat{b}}
\def\hc{\hat{c}}
\def\hd{\hat{d}}
\def\hi{\hat{i}}
\def\hj{\hat{j}}
\def\hk{\hat{k}}
\def\ho{\hat{0}}
\def\h1{\hat{1}}
\def\h2{\hat{2}}
\def\h3{\hat{3}}
\def\h4{\hat{4}}

\def\a{\alpha}
\def\b{\beta}
\def\c{\gamma}
\def\d{\delta}
\def\db{\bar{\delta}}
\def\l{\lambda}
\def\u{\mu}
\def\v{\nu}
\def\th{\theta}
\def\ph{\phi}
\def\pd#1{\partial_{#1}}
\def\O{{\mathcal{O}}}
\def\D{{\mathcal{D}}}

\begin{document}
\bibliographystyle{apsrev}

\preprint{}

\title{Spin-induced non-geodesic motion, gyroscopic precession, Wigner rotation and
EPR correlations of massive spin-$\half$ particles
in a gravitational field}



\author{P.M. Alsing$^\dagger$}
\email{alsing@hpc.unm.edu}
\thanks{Corresponding author}
\author{G.J. Stephenson Jr}
\email{gjs@swcp.com}
\author{Patrick Kilian}
\email{pkilian@physik.uni-wuerzburg.de}
\affiliation{$^*$Air Force Research Laboratory, Space Vehicles Directorate \\
3550 Aberdeen Ave, SE, Kirtland AFB, New Mexico, 87117-5776}
\affiliation{$^{\dagger}$Department of Physics and Astronomy, University of New
Mexico, Albuquerque, NM 87131}
\affiliation{\\$^\ddagger$Bayrische Julius-Maximilians Universit\"{a}t W\"{u}rzburg, Germany}

\date{\today}

\begin{abstract}
We investigate in a covariant manner the spin-induced non-geodesic motion of massive spin-$\half$ particles in an arbitrary
gravitational field for trajectories that are initially geodesic when spin is ignored.
Using the WKB approximation for the wave function in curved spacetime, we compute the ${\mathcal{O}}(\hbar)$ correction to
the Wigner rotation of the spin-$\half$ particle, whose ${\mathcal{O}}(1)$ contribution is zero on timelike geodesics.
We develop conditions for the motion of observers in which the Wigner rotation is null.
For the spherically symmetric Schwarzschild metric, we consider
specific examples of particle motion in the equatorial plane for
(i)  circular orbits  and
(ii) radially infalling trajectories. For the former case we
consider the entanglement for a perfectly anti-correlated EPR entangled pair of spins as the separate qubits traverse the
circular orbit in same direction.
\end{abstract}
\pacs{}

\maketitle





%
%


\section{Introduction}
Entanglement is an important resource for many applications in quantum information science (QIS) including
teleportation, quantum computation, and quantum communication.
Recently, there has been a growing interest in understanding entanglement in quantum information science
beyond the confines of its non-relativistic quantum mechanical origins. A excellent recent review can be found in
Peres and Terno \cite{peres_terno} and references therein. Initial studies concerned the behavior of
quantum states, both single particle and bipartite entangled states under the action of Lorentz transformations (LT)
\cite{czachor}, which transforms between different inertial (constant velocity, zero acceleration) observers.
When quantum mechanics is merged with special relativity (SR) in the form of quantum field
theory (QFT), the state of a particle is labeled not only by its spin (or helicity)
(as in non-relativistic quantum mechanics (NRQM)),
but also by its momentum. These two quantities represent the Casimir invariants of operators which
commute with the Poincare transformations (LTs plus translations) which underly the symmetries of
flat (Minkowski) spacetime, where SR applies.

Peres \tit{et al} \cite{peres_scudo_terno} were among the first to point out that even
for a single particle, this could lead to an observer dependent change of the (von Neumann) entropy of the reduced
spin density matrix when the momentum is traced out after the action of a LT.
Alsing and Milburn \cite{alsing_milburn} investigated the transformation of maximally entangled bipartite
Bell states composed of pure momentum eigenstates, and showed that while initially colinear spin
and momentum directions are transformed under a LT to non-colinear directions, the overall entanglement
is preserved. The amount by which the spin of a massive particle is rotated is given by the momentum dependent
Wigner rotation angle, discussed in Alsing and Milburn and reviewed in the main body of the text below.
Subsequent papers \cite{epr_papers} explicitly pointed out the implied consequence of the
previous work that there would be an apparent decrease in the magnitude of a Bell measurement if
the measurement was made along the Lorentz transformed momentum direction, but the original, untransformed value would be
obtained if the measurement was made along the transformed spin direction. Gingrich, Bergou and Adami \cite{adami_spin}
pointed out that since the Wigner angle is momentum dependent (since in general, a LT changes the magnitude
of the particle's momentum), a wavepacket state composed of an integration of single particle states
over a momentum distribution, would have the spin of each component state transformed differently,
according to the value of its momentum. They investigated a wavepacket state for a bipartite state of
two spin $\half$ particles and showed that, due to spin-momentum entanglement,
the reduced two particle spin density matrix (formed by tracing out the momentum) had a Wootters'
concurrence \cite{wootters} (an exact measure of two qubit entanglement) that depended on the inertial frame from which
the state was observed (i.e. upon the LT considered). Similar considerations for the transformation
of photon states under LTs were also considered by several authors \cite{alsing_milburn,adami_photon,rembielinski}.

A step towards more general types of motion was considered by several authors who considered entanglement
for constantly accelerated observers in flat spacetime \cite{unruh_papers}. It has been well
known that these \tit{Rindler} observers measure a thermal flux of particles (Bose-Einstein for bosons,
Fermi-Dirac for fermions) as they move through the flat spacetime Minkowski vacuum,
at a temperature that is proportional to the observer's acceleration (the Unruh effect, which
is the flat spacetime analogue of the Hawking effect of black hole evaporation).
These author's investigated
the fidelity of teleportation and other entanglement measures for maximally entangled states of
both spin zero and spin $\half$ particles.

An important step forward in the evolution of these relativistic investigations was made by Terashima and Ueda \cite{ueda} who
investigated the transformation of single particle and entangled states under arbitrary states of
motion (acceleration), where general relativity (GR) applies. Essentially, the global inertial frames
of SR (zero acceleration) are now replaced by local reference frames, tangent to the curved spacetime (CST)
at the point $x$, for an arbitrary accelerating
observer at the spacetime point $x$. This observer is described in terms of a \tit{tetrad}, or four
4-vector axes, three of which describe the spatial axes of the observer's local laboratory at $x$ and
one temporal axis which governs the local rate at which his clock ticks (the observer's \tit{proper time}).
Inside the observer's local laboratory at the point $x$ SR holds, which is an embodiment of Einstien's
Equivalence Prinicple (EP).
An observer makes measurements of a particle that passes through his local laboratory at $x$ by projecting
the particle's momentum onto the the four axes of his tetrad. Terashima and Ueda showed that
as the particle moves infinitesimally from $x\to x'$ in the CST, the spin of a particle
is transformed by a \tit{local Lorentz transformation} (LLT), and correspondingly by a local Wigner rotation
of its spin. In general, a LLT transforms between different observers in arbitrary states of motion,
all instantaneously at the CST point $x$ (e.g. stationary, freely falling, circular geodesic, or
arbitrary acceleration), as will be detailed in Section II.

In this paper we show that a consequence of the above considerations is that the Wigner rotation is measured
to be null in the non-rotating, instantaneous rest frame of the accelerating particle, the so called
\tit{Fermi-Walker frame} (FWF). In any other reference frame, the observer would detect a non-zero Wigner
rotation angle that is dependent upon his particular state of motion.
 If the particle is undergoing force free \tit{geodesic} motion (zero acceleration)
the FWF reduces to the freely falling frame (FFF) in which all four axes of the tetrad are parallel transported
along the particles 4-velocity (the tangent to the particle's geodesic, which is equal to the temporal axis
of the tetrad). It is a postulate of GR that the force free motion of particles (massive or massless) follow
geodesics. This is true if the spin of the particle is assumed to be zero. However, for particles with spin,
even classically spinning particles \cite{carmeli}, the spin of the particle couples to the curvature
and leads to non-geodesic (accelerated) motion. In this work, we consider quantum spin half particles
whose orbits are initially geodesic, if spin is ignored, and consider the $\O(\hbar)$ corrections to
their motion when the particle's motion is defined by its quantum mechanical Dirac current \cite{audretsch,thorne_bhmp}.
We develop the $\O(\hbar)$ Wigner angle for such particle motion and investigate the implications of entanglement
of Bell states in CST. A companion article \cite{alsing_stephenson_photon} explores these considerations
for photon states and the effects of the local Wigner rotation in CST on entangled photon states.

This paper is organized as follows. In Section II we review freely falling frames and Fermi-Walker transported
frames in general relativity. In Section III we review the Wigner rotation and the transformation of
massive positive energy, single particle states in flat spacetime, while in Section IV we generalize this
to curved spacetime and review the work of Terashima and Ueda. In the Appendix A we provided a detailed
derivation of the formula for the local Wigner rotation angle, not provided in the previous work \cite{ueda}.
In section V we develop the Dirac equation for spin $\half$ particles in curved spacetime, and
in Section VI derive a WKB approximation to its solution. In Section VII we consider the $\O(\hbar)$
velocity and acceleration corrections to the initially circular geodesic motion of particles
in the spherically symmetric Schwarzschild spacetime (derived in detail in Appendix B),
when spin is ignored, and discuss the consequences for entangled
Bell states on two neighboring, infinitesimally close
circular orbits. We also discuss the spin-momentum entanglement of wavepacket states, and show that in
CST, the Wigner rotation is also dependent upon the initial orientation of the particle's spin in
its reference frame.
In Section VIII we extend the discussion to the case of radially infalling geodesic motion.
In Section IX we present a brief summary, and our conclusions.

\section{Freely falling frames and Fermi-Walker transported frames in General Relativity}
In Netwonian mechanics an inertial frame $S$, i.e. the laboratory from which an observer can make measurements,
is defined as follows: (1) chose as the origin of $S$ a free particle, for all time and (2) at one instant of time,
chose three mutually orthogonal spatial axes defined by the orientations of three perpendicular gyroscopes. At later times,
continue to define the spatial axes by the directions of the three orthogonal gyroscopes. Equivalently,
parallel transport (with no rotation) the initial directions of the gyroscopes along the straight line trajectory of the
free particle. The directions of the parallel transported spatial axes can be used to define
Cartesian coordinates $(x,y,z)$ with respect to the origin. All other reference frames (other laboratories)
that move with constant velocity with respect to this inertial frame, are also inertial frames themselves.
In special relativity (SR), an inertial frame is defined in exactly the same fashion,
except now the universal time of Newtonian mechanics has to be abandoned. That is, a particular inertial frame $S$
defines four Cartesian coordinates $(t,x,y,z)$, while a different inertial frame $S'$, travelling with constant velocity
with respect to $S$, defines the coordinates $(t',x',y',z')$, which are related to the coordinates of $S$ by
the usual Lorentz transformations.

\subsection{Reference frames in general relativity}
In general relativity (GR) the global inertial frames of flat (Minkowski) spacetime have to be abandoned for a description in
terms of \tit{local inertial frames} (LIF), which are the SR inertial frames valid now for only a limited range of
the coordinates, both spatial and temporal. This is just a statement of Einstein's equivalence principle, that an
arbitrary spacetime is \tit{locally} flat. To generalize the global inertial frames of SR to LIFs in an arbitrary
curved spacetime, one introduces four spacetime dependent, mutually orthogonal axes $\bfe{a}(x)$,
where the hatted index $\ha$ labels the four local axes $\ha=(\ho,\hat{1},\hat{2},\hat{3})$,
such that $\bfe{a}(x)\cdot\bfe{b}(x) = \mbf{\eta}$. Here, $\mbf{\eta} =$ diagonal$(1,-1,-1,-1)$ is the
flat spacetime metric of SR. In component form we have
\be{1}
   g_{\u\v}(x)\, \e{a}{\u}(x) \, \e{b}{\v}(x) = \eta_{\ha \hb}, \qquad \u,\v = (0,1,2,3).
\ee
In the above, $g_{\u\v}(x)$ is the metric of the curved spacetime with line element $ds^2 = g_{\a\b}(x)\,dx^\a\,dx^\b$.
We will use units in which the speed of light is set to unity, $c=1$. With our chosen metric signature $(+,-,-,-)$
we can define the proper time $\tau$ as $d\tau^2 = ds^2$. Free massive particles ($m\ne 0)$ follow
timelike geodesics ($ds^2>0$) in the curved spacetime, while massless particles ($m=0$) follow null or lightlike
geodesics ($ds^2 = 0$). In this work, we will use the term particle to mean massive objects (electrons, protons, etc\ldots)
and refer explicitly to massless objects (e.g. photons) when needed.

As a comment on notation, in a specific set of coordinates, e.g. $x^\a = (t,r,\th,\ph)$,
$\bfe{0}(x)=\Big( \e{0}{t}(x), \e{0}{r}(x), \e{0}{\th}(x), \e{0}{\ph}(x)\Big)$
are the components of the timelike axis
in the coordinate basis defined by $x^\a$
( i.e. $\mbf{e}_{\a}(x)$ such that $\mbf{e}_{\a}(x)\cdot\mbf{e}_{\b}(x) = g_{\a\b}(x)\,$ )
and $\bfe{i}(x)=\Big(\e{i}{t}(x)\e{i}{r}(x), \e{i}{\th}(x), \e{i}{\ph}(x)\Big)$
are the components of the spacelike axis $\bfe{i}$, where $i = (1,2,3)$. We will denote by $\mathbf{e}(x)$
the collection of all four axes into a matrix for which the $\ha$th row is $\bfe{a}(x)$.

The components $\e{a}{\u}(x)$ are called a \tit{tetrad} or \tit{vierbien} (four-legs) \cite{tetrads}.
Note that the relationship between the orthonormal basis $\bfe{a}(x)$ and the coordinate basis $\mbf{e}_{\a}(x)$
(with components defined by $(\mbf{e}_{\a})^\b = \delta_\a^{\spp \b}$  using the coordinates $x^\alpha$),
 is given by $\bfe{a}(x) = \e{a}{\a}(x)\,\mbf{e}_{\a}(x)$.
In the following we will need
the inverse matrix of tetrads $\mathbf{e}^{-1}(x)$ with components denoted by $\inve{\u}{a}(x)$. In addition, we will also
need the transpose of this matrix $\mathbf{e}^{-1 T}(x)$, where $T$ denotes transpose, with components
given by $\inveT{a}{\u}(x)$. The inverse tetrads satisfy the dual to \Eq{1} i.e.
\be{2}
 g^{\u\v}(x) \, \inve{\u}{a}(x) \, \inve{\v}{b}(x) = \eta^{\ha \hb}
\ee
where $\eta^{\ha \hb}$ and $g^{\u\v}(x)$ are the inverse flat and curved spacetime metrics, respectively.
Note that in matrix from we can write \Eq{1} as $\mathbf{e}\cdot\mbf{g}\cdot\mathbf{e}^T = \mbf{\eta}$ and
\Eq{2} as $\mbf{e}^{-1 T}\cdot\mbf{g}^{-1}\cdot\mbf{e}^{-1} = \mbf{\eta}$, where in the last expression
we have used $\mbf{\eta}^{-1} = \mbf{\eta}$. The matrix $\mathbf{e}^{-1}$ is the inverse of the tetrad $\mathbf{e}$ as
can be seen in component form by
\be{3}
 \e{a}{\u}(x) \, \inve{\u}{b}(x) = \delta_{a}^{\spp b}, \qquad \inve{\u}{a}(x) \, \e{a}{\v}(x) = \delta_{\u}^{\spp \v}.
\ee
We denote Greek indices $\{\u,\v,\ldots\}$ as \tit{world} indices in the arbitrary spacetime and (hatted) Latin indices
$\{\hat{a},\hat{b},\ldots\}$ as \tit{local Lorentz} indices in the observer's LIF.  World indices are raised and lowered
with the curved spacetime metric $g_{\a\b}(x)$ and Latin indices are raised and lowered
with the local flat Minkowski metric $\eta_{\ha\hb}$.
Thus, the transpose inverse components $\inveT{a}{\u}(x)$ are related to the tetrad components $\e{a}{\u}(x)$ by
$\inveT{a}{\u}(x) = \eta^{\hat{a} \hat{b}} \, \e{b}{\v}(x) \, g_{\v\u}(x)$.

A vector $\mbf{V}$ is a geometric object which can be decomposed in either the coordinate basis $\{\mbf{e}_\a(x)\}$ or
the local orthonormal basis $\{\mathbf{e}_{\ha}(x)\}$, i.e.
$\mbf{V} = V^\a(x) \, \mbf{e}_\a(x) = V^{\ha}(x)\,\mathbf{e}_{\ha}(x)$.
The utility of the tetrad and inverse tetrad is that the observer can obtain the local values $V^a(x)$
of the components of a  world vector $V^\a(x)$ by
projecting the world vector onto the observer's four local axes
\bea{4}
V^{\ha}(x) &=&  V^\a(x) \, \inve{\a}{a}(x), \qquad  V^\a(x) = V^{\ha}(x) \, \e{a}{\a}(x),\\
           &=&  \inveT{a}{\a}(x)\,V^\a(x). \nonumber
\eea
For a general tensorial object $T^{\a\b}_{\sp\sp\sp\sp\gamma}(x)$, we find its LIF components $T^{\ha \hb}_{\sp\sp\sp \hc}(x)$
by a similar projection onto the
observer's local axes
\be{5}
T^{\ha \hb}_{\sp\sp\sp \hc}(x) = T^{\a\b}_{\sp\sp\sp\sp\gamma} \,\inve{\a}{a}(x)\,\inve{\b}{b}(x) \, \e{c}{\gamma}(x).
\ee

By using this set of orthonormal axes (basis vectors), the observer has made the metric of his laboratory locally
flat, $\bfe{a}(x)\cdot\bfe{b}(x) = \eta_{\ha\hb}$. The observer can subsequently construct coordinates in his laboratory
such that the derivative of the metric $g_{\a\b}(x)$ vanishes all along the the geodesic trajectory of the
origin of his laboratory (Riemann normal coordinates). These coordinates are valid only if the
observer's laboratory is sufficiently ``small" spatially and measurements are made over ``short enough" times,
otherwise, these coordinates lines can cross each other and thus become invalid for making observations.
Such a reference frame, in which $g_{\a\b}(x_\O)\to\eta_{a b}$ and $\pd{\u} g_{\a\b}(x_\O) = 0$ where
$x^\a_\O$ are the coordinates of the origin of the observer's laboratory along its geodesic trajectory in
the curved spacetime, is called a \tit{freely falling frame} (FFF).

The above FFF is the local analogue of the inertial frame of SR. For the motion of free particles, i.e. geodesics, it
is the most ``natural" frame from which the observer can make measurements. However, in general this is not
the only way to define the observer's local laboratory. For example, the origin of a stationary observer's laboratory
which sits at fixed spatial coordinates would in general experience an acceleration
and possible spatial rotations. All local laboratories can
be related to each other by spacetime dependent local Lorentz transformations (LLT) relating their choice of orthonormal bases i.e.
$\mathbf{e}'_{a}(x) = \Lambda_{\ha}^{\spp \hb}(x)\, \bfe{b}(x)$. The LLTs are independent of the general coordinate
transformations (GCT) $\partial x'^\a(x)/\partial x^\b$ that can be made in the curved spacetime, that
relate quantities (vectors, tensors, etc\ldots) in the same spacetime described in the new coordinates $x'^\a(x)$.
Thus, the tetrad
components $\e{a}{\a}(x)$ transforms as a contravariant world vector (index $\a$) under general coordinate transformations, and
as a covariant local Lorentz vector (index $\ha$) under LLTs. The local metric $\eta_{\ha\hb}$ transforms as a
symmetric covariant local Lorentz tensor of rank two under LLTs, but as a scalar under GCTs. The reverse is true
for the world metric $g_{\a\b}(x)$ which transforms as a scalar under LLTs and a covariant tensor of rank two under GCTs.
As a computational consequence of the freedom to make arbitrary LLTs, one can always perform computations in GR
with the components of vectors (tensors) referred to a coordinate basis $\mbf{V}(x) = V^\a(x)\,\mbf{e}_\a(x)$, and then transform
to the components $V^{\ha}(x)$ with respect to an orthonormal basis $\bfe{a}(x)$ by a LLT as in \Eq{4} and \Eq{5}.

If $\mbf{u}(x) = u^\a(x) \, \mbf{e}_\a(x)$ is the 4-velocity of a free particle with
components $u^\a(x) = dx^\a/d\tau$ in a coordinate basis, the condition for force-free or geodesic motion is
\be{6}
 \nabla_{\mathbf{u}} \, u^\a(x) \equiv u^\b(x) \, \nabla_\b \, u^\a(x)
 =  u^\b(x) \left( \pd{\b} \, u^\a(x) + \Gamma^{\a}_{\sp \l\b}(x)\,u^\l(x)  \right)= 0.
\ee
In general, $\nabla_{\mathbf{u}} \, \mbf{V}(x) =  \big(\nabla_{\mathbf{u}} V^\a(x)\big)\, \mbf{e}_\a(x)$
is the total derivative (or directional derivative) of the
the vector $\mbf{V}(x)$ along the geodesic with tangent $\mbf{u}$, which generalizes to curved spacetime
the concept of the derivative $d/d\tau$ along a curve.
$\nabla_\b$ is called the covariant derivative
and, when one is concerned solely with the components of a vector, $\nabla_\b V^\a(x)$ is referred to as the
covariant derivative of the vector $V^\a(x)$, which is commonly denoted as $V^\a_{\sph\spp ;\b}(x)$.
In the above, $\Gamma^{\a}_{\sp \l\b}(x)$ is the usual metric (Christoffel) affine connection
defined with respect to a coordinate basis $\mbf{e}_\a(x)$
from $\nabla_{\b}\mbf{e}_\a(x) = \Gamma^{\l}_{\sp \a\b}(x)\,\mbf{e}_\l(x)$,
and computable from the metric via the metric compatibility condition $\nabla_{\u}\,g_{\a\b}(x)=0$,
\be{7}
\Gamma^\a_{\sp \u\v}(x) = \half\, g^{\a\b}\,\Big( \pd{\u}\, g_{\b\v}(x) + \pd{\v}\, g_{\b\u}(x) - \pd{\b}\, g_{\u\v}(x) \Big).
\ee
The Christoffel connection indicates how the tetrad twists and turns as it moves from $x\to x'$, by
specifying the rule for parallel transport of tensorial objects in the CST.
Equation~(\ref{6}) states the directional derivative
of the 4-velocity $\u(x)$ along itself is zero, which is the geometrical statement that the 4-velocity is parallel transported along itself.
More physically, the components of the 4-acceleration experienced by a particle along an arbitrary trajectory with a
general 4-velocity $\mbf{u}$ in a coordinate/orthonormal basis is given by
\be{8}
a^\a(x) = u^\b(x) \, \nabla_\b \, u^\a(x), \qquad a^{\ha}(x) = a^\a(x) \inve{\a}{a}(x).
\ee
This is the external, non-gravitational acceleration that would have to be applied to keep the particle on
its trajectory.
Thus, \Eq{6} is a statement that the origin of the observer's laboratory experiences no acceleration along its
geodesic trajectory. To construct the FFF of the observer's laboratory, we define the orthonormal basis as follows:
(1) we choose the temporal basis vector $\bfe{0}(x)$ to be the observer's 4-velocity $\mbf{u}(x)$ and (2) require that
the remaining orthogonal spatial basis vectors $\bfe{i}(x)$ be parallel transported along the geodesic, i.e.
\be{9}
\trm{FFF condition:}\qquad \bfe{0}(x) \equiv \mbf{u}(x), \quad \nabla_{\mathbf{u}} \, \bfe{a}(x)
= u^\beta(x)\nabla_\beta \, \bfe{a}(x)=0 \qquad \ha=(\ho,\hat{1},\hat{2},\hat{3}).
\ee

\subsection{Fermi-Walker transported frames in general relativity}
So far the issue of the non-rotation of the laboratory axes along its trajectory in a curved spacetime has not been addressed.
For  general non-geodesic motion with 4-velocity $\mbf{u}$ (from now on dropping the indication of the spacetime dependence
$(x)$ unless explicitly needed)
a particle will experience a 4-acceleration $\mbf{a}$.
Note that the 4-velocity of a massive particle is normalized to unity, $\mbf{u}\cdot\mbf{u} = 1$, which follows
directly from the expression for the metric $ds^2 = d\tau^2 = g_{\a\b}(x)\,dx^\a\,dx^\b$ and the definition
of the 4-velocity as $\mbf{u} = d\mbf{x}/d\tau$. Consequently, by taking the covariant derivative of
both sides of this normalization equation one finds that the 4-acceleration $\mbf{a} = \nabla_{\mathbf{u}}\,\mbf{u}$
is orthogonal to the 4-velocity, $\mbf{a}\cdot\mbf{u}=0$.
If we use this particle as the origin of an observer's laboratory frame, we require that the tetrads carried by
the observer ``not rotate." Some care has to be taken in curved spacetime when clarifying the precise definition of the
concept of the non-rotation of the tetrad, since, as the particle progresses along its trajectory, its 4-velocity will change, which can
be considered as a rotation (LLT) in the instantaneous frame defined by $\mbf{u}$ and $\mbf{a}$. Non-rotation then
means we accept this inevitable rotation, but require that any spacelike vector $\mbf{w}$ orthogonal to both
$\mbf{u}$ and $\mbf{a}$ undergo no additional spatial rotation. For a general vector $\mbf{V}$ this last
requirement is ensured by the transport law
\be{10}
\nabla_{\mathbf{u}}\,\mbf{V}
= (\mbf{V}\cdot\mbf{u})\,\mbf{a} - (\mbf{V}\cdot\mbf{a})\,\mbf{u} \equiv \mbf{\Omega}(\mbf{V}).
\ee
The rotation tensor $\mbf{\Omega}$ accomplishes the required task as evidenced from the following special case:
(i) $\mbf{\Omega}(\mbf{u})=\mbf{a}$, which returns the definition of the 4-acceleration, and (ii) for a spatial vector
$\mbf{w}$ orthogonal to both $\mbf{u}$ and $\mbf{a}$, $\mbf{\Omega}(\mbf{w})=0$, which reduces to the
statement that $\mbf{w}$ is parallel propagated along the trajectory.
A vector evolving according to the rule in \Eq{10} is said to be \tit{Fermi-Walker} (FW) \tit{transported}
along the the particle's trajectory with tangent $\mbf{u}$.
A tetrad $\mathbf{e}(x)$ with each axis satisfying \Eq{10} is called a \tit{Fermi-Walker frame} (FWF), and
represents the instantaneous, non-rotating rest frame of an accelerating particle.
Note that for geodesic motion, in which $\mbf{a}=0$,
FW transport reduces to parallel transport. In later sections, we will see explicit examples of parallel and
FW transported motion of the tetrad defining the observer's local coordinate frame.

\section{Wigner Rotation}
\subsection{Flat Spacetime}
In flat (Minkowski) spacetime, the positive energy, single particle state of a massive  particle forms
a spinor representation of the inhomogeneous Lorentz (Poincare) group \cite{weinberg_qft1}.
These states denoted by $\ket{\vec{p},\sigma}$,
are labelled by their spatial momentum $\vec{p}$ (where $p^0 = E = \sqrt{\vec{p}^{\sph 2} + m^2}$))
and the component $\sigma$ of spin along a quantization axis in its rest frame (typically taken to be along the third spatial direction).
Under a Lorentz transformation $\mbf{\Lambda}$ the one-particle state transforms under the unitrary transformation
$U(\Lambda)$ as
\be{11}
U(\Lambda) \ket{\vec{p},\sigma} = \sum_{\sigma'} \,
D^{(j)}_{\sigma'\sigma}(W(\Lambda,\vec{p}))\,\ket{\overrightarrow{\Lambda p},\sigma'},
\ee
where $j$ is the spin of the particle, the summation is over $\sigma' = (-j,-j+1,\ldots,j)$ and
$\overrightarrow{\Lambda p}$ are the spatial components of the Lorentz transformed 4-momentum, i.e.
$\vec{p}^{\sph '}$ where $p^{'\u} = \Lambda^\u_{\spp \v} \, p^\v$. In this work we will be
primarily concerned with spin-$\half$ Dirac particles $(j=\half)$.
In \Eq{11}, $D^{j}_{\sigma'\sigma}(W(\Lambda,\vec{p}))$ is a $(2 j+1)\times (2 j+1)$ matrix
spinor representation of the rotation group $SU(2)$, and $W(\Lambda,\vec{p})$ is called the Wigner rotation angle.
The explicit form of the Wigner rotation in matrix form is given by
\be{12}
 \mbf{W}(\Lambda,\vec{p}) =  \mbf{L}^{-1}(\Lambda p) \cdot \mbf{\Lambda} \cdot \mbf{L}(p),
\ee
where $\mbf{L}(p)$ is a \tit{standard boost}  taking the standard rest frame 4-momentum $\mbf{k}\equiv (m,0,0,0)$
to an arbitrary 4-momentum $\mbf{p}$, $\mbf{\Lambda}$ is an arbitrary LT taking
$\mbf{p}\to \mbf{\Lambda}\cdot \mbf{p} \equiv \mbf{\Lambda p}$, and
$\mbf{L}^{-1}(\Lambda p)$ is an inverse standard boost taking the final 4-momentum $\mbf{\Lambda p}$ back to
the particle's rest frame. Because of the form of the standard rest 4-momentum $\mbf{k}$, this final
rest momentum $\mbf{k}^{\sph '}$ can at most be an spatial rotation of the initial standard 4-momentum $\mbf{k}$, i.e.
$ \mbf{k}^{\sph '} = \mbf{W}(\Lambda,\vec{p}) \cdot \mbf{k}$. The rotation group $O(3)$ is then said to form
(Wigner's) \tit{little group} for massive particles, i.e. the invariance group of the particle's rest 4-momentum.
The explicit form of the standard boost is given by \cite{weinberg_qft1}
\bea{13}
%
 L^0_{\spp 0} &=& \gamma = \frac{p^{0}}{m}\no
 L^i_{\spp 0} &=& \frac{p^{i}}{m}, \quad L^0_{\spp i} = -\frac{p_{i}}{m}, \no
 L^i_{\sp j} &=& \delta^i_{\sp j} - (\gamma-1)\, \frac{p^i p_j}{|\vec{p}\,|^2}, \qquad i,j = (1,2,3),
\eea
where $\gamma = p^{0}/m = E/m \equiv e$ is the particles energy per unit rest mass. Note that
for the flat spacetime metric $\eta_{\a\b} =$diag$(1,-1,-1,-1)$, $p_0 = p^0$ and $p_i = -p^i$.

\section{Curved Spacetime}
In curved spacetime, essentially everything above goes through unchanged except for
the important fact that single particle states now form a \tit{local} representation of the inhomogeneous Lorentz group
at each spacetime point $x$. Thus, a single particle state is now represented as
$\ket{p^{\hi}(x),\sigma}\;$ $i=(1,2,3)$ where $p^{\hi}(x)$ are the spatial components of the particle's 4-momentum
$\mbf{p} = p^{\ha}(x)\, \bfe{a}(x)$ in the local orthonormal basis $\{\bfe{a}(x)\}$. Since
the particle's local 4-momentum transforms under LLTs $\mbf{\Lambda}(x)$,
the single-particle state transforms unitarily via the local version of \Eq{11}
\be{14}
U(\Lambda(x)) \ket{p^{\hi}(x),\sigma} = \sum_{\sigma'} \,
D^{(j)}_{\sigma'\sigma}(W(x))\,\ket{(\Lambda p)^{\sph\hi}(x),\sigma'},
\ee
where $\mbf{W}(x)$ is the local Wigner rotation given by the local version of \Eq{12},
\be{15}
 \mbf{W}(x) \equiv  \mbf{L}^{-1}(\Lambda p(x)) \cdot \mbf{\Lambda}(x) \cdot \mbf{L}(p(x)).
\ee

\begin{figure}[h]
\includegraphics[width=3.75in,height=2.75in]{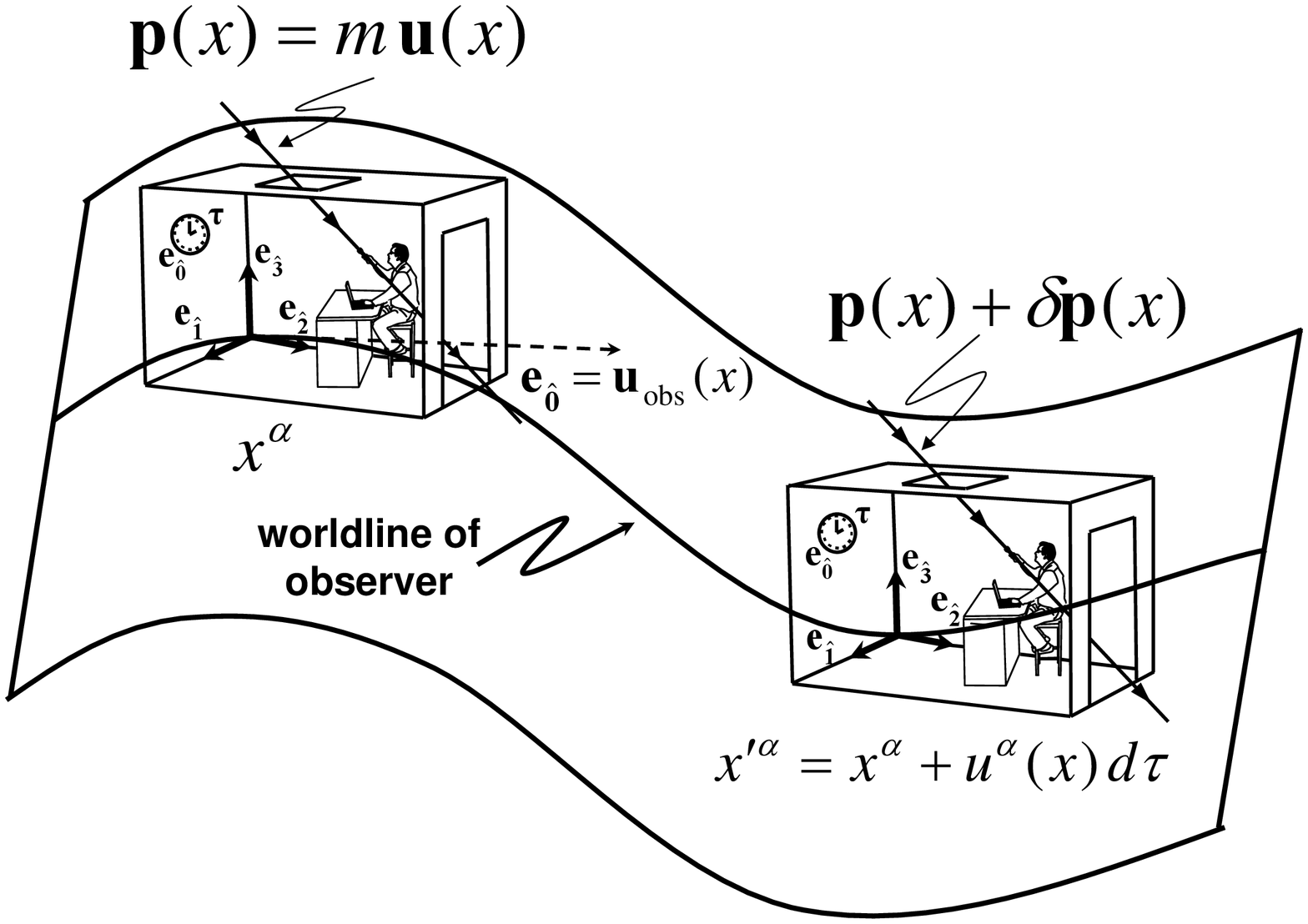}
\caption{The observer's \tit{local laboratory} (small box with man) at the curved spacetime point $x$, defined
by the orthonormal tetrad $\bfe{a}(x)$. The three spatial axes $\bfe{i}(x)$, $\hi = (\hat{1},\hat{2},\hat{3})$
are located at the origin of the observer's laboratory, while $\bfe{0}(x) = \mathbf{u}_{obs}(x)$ is
the temporal axis, defined as his 4-velocity, or the tangent to his geodesic trajectory.
A particle of 4-momentum $\mathbf{p}(x) = m\,\mathbf{u}(x)$ and \tit{world} components $p^{\a}(x)$ passes through the observer's
local laboratory. The observer measures the \tit{local} components $p^{\ha}(x)$ of the particle
by projecting $\mathbf{p}(x)$ onto the four tetrad axes, $p^{\ha}(x) = \inveT{a}{\a}(x)\,p^\a(x)$.
At a small proper time later $d\tau$, the particle has moved from $x^\a\to x'^\a = x^\a + u^\a(x)\,d\tau$,
which is measured by the observer in his local laboratory at the spacetime point $x'^\a$.
}
\label{fig1}
\end{figure}

Following Terashima and Ueda \cite{ueda}, let us consider how the spin changes
as we move from one spacetime point in curved spacetime to another along an arbitrary timelike trajectory. Let our
particle initially be at a spacetime point with coordinates $\mbf{x}$ and
4-momentum $\mbf{p}(x) = m \mbf{u}(x)$. At a small proper time $d\tau$ later the particle
has moved along its trajectory with tangent $\mbf{u}$ to the point
with coordinates $\mbf{x}' = \mbf{x} + \mbf{u}(x)\,d\tau$ and new 4-momentum
$\mbf{p}(x) + \delta\mbf{p}(x)$, illustrated in \Fig{fig1} \cite{LLO_comment}.
Since the spin of the particle is defined locally with
respect to the observer's reference frame, defined by the tetrad that is carried
along with him at the laboratory's origin, we are interested in the momentum change $\delta\mbf{p}(x)$
relative the orthogonal basis vectors, i.e. $\delta\mbf{p}(x) = \delta p^{\ha}(x) \, \bfe{a}(x)$.
For small $d\tau$ we expect $\delta p^{\ha}(x)$ to be proportional $p^{\ha}(x)$ and to $d\tau$ so
we seek an expression of the local 4-momentum change in the form of
\be{16}
\delta p^{\ha}(x) = \lambda^{\ha}_{\spp \hb}(x) \,p^{\hb}(x)\, d\tau.
\ee
where $\lambda^{\ha}_{\sph \hb}(x)$ is an infinitesimal LLT,
\be{17}
\Lambda^{\ha}_{\sph \hb}(x) = \delta^{\ha}_{\spp \hb} + \lambda^{\ha}_{\sph \hb}(x)\,d\tau.
\ee

From the definition of the local 4-momentum in the observer's reference frame as
a projection of the world 4-momentum onto the local axes
$p^{\ha}(x) = \inveT{a}{\u}(x)\,p^{\u}(x)$ we have
\be{18}
\delta p^{\ha}(x) = \inveT{a}{\u}(x)\; \delta p^{\u}(x) + \delta\inveT{a}{\u}(x)\; p^{\u}(x).
\ee
Equation~(\ref{18}) contains two terms; the first $\delta p^{\u}(x)$ is the change of the
world 4-momentum components $p^{\u}(x)$
as the particles moves from $x^\u \to x^{'\u}$ in the underlying curved spacetime, and the
second $\delta\inveT{a}{\u}(x)$ is is the change in the tetrads components (here, the
inverse transpose components) $\inveT{a}{\u}(x)$
which are used to project the world 4-momentum components onto the observer's local laboratory axes,
 $p^{\u}(x) \to p^{\ha}(x)$. The first term is simply given by
\be{19}
\delta p^{\u}(x) = \nabla_{\mathbf{u}} \, p^{\u}(x) \, d\tau
= u^\v(x) \, \nabla_{\v} \,  p^{\u}(x) \, d\tau = m \, a^\u(x) d\tau,
\ee
where we have used the definition of the 4-momentum in terms of the 4-velocity $\mbf{p}(x) = m \mbf{u}(x)$
and the definition of the acceleration in \Eq{8}. Since $\mbf{p}\cdot\mbf{p} = (m c)^2 = m^2$ we can
write \Eq{19} as
\bea{20}
\delta p^{\u}(x) &=& \frac{1}{m} \, \left[ a^\u(x) \, p_\v(x)- p^\u(x) \, a_\v(x)   \right] \, p^\v(x) \, d\tau, \\
&\equiv&  \frac{1}{m} [\mbf{\Omega}(\mbf{p})]^\u(x) \, d\tau
= \frac{1}{m} \, \Omega^{\u}_{\spp \v}(x) \, p^\v(x) \, d\tau, \nonumber
\eea
where $\mbf{\Omega}$ is the ``non-rotation" matrix (i.e ensures rotation only in the $\mbf{u}$-$\mbf{a}$ plane)
on the right hand side of the Fermi-Walker transport equation in \Eq{10}.

For the second term in \Eq{18}, we note that inverse transpose of the tetrads define a set of 1-forms defined
by $\mathbf{e}^{\ha}(x) = \inveT{a}{\u}(x)\, dx^\u$ dual to the orthonormal basis vectors $\bfe{a}(x)$.
The covariant derivative of an arbitrary set of basis vectors (orthonormal or not) is given
by $\nabla_\v \, \bfe{a}(x) = \omega_{\v \sp\sp \ha}^{\spp \hb}(x) \, \bfe{b}(x)$, which generalizes
the Christoffel  connection discussed just before \Eq{7}, defined in terms of a coordinate $\mbf{e}_\a(x)$
basis by $\nabla_{\b}\mbf{e}_\a(x) = \Gamma^{\l}_{\sp \a\b}(x)\,\mbf{e}_\l(x)$. Since the 1-forms
$\mathbf{e}^{\ha}(x)$ are dual to the orthonormal basis vectors $\bfe{a}(x)$ in the sense that
$\mathbf{e}^{\ha}(x)\cdot \bfe{b}(x)=\delta^{\ha}_{\sp \hb}$, the covariant derivative of
the 1-forms (using $\nabla_\v \delta^{\ha}_{\spp \hb} = 0$) is given by
$\nabla_\v \mathbf{e}^{\ha}(x) =  -\omega_{\v \sp\sp \ha}^{\spp \hb}(x)\,\mathbf{e}^{\ha}(x)$.
Expanding this out in terms of coordinate 1-forms $dx^\u$
yields $\nabla_\v \inveT{a}{\u}(x) =  -\omega_{\v \sp\sp \hb}^{\spp \ha}(x)\,\inveT{a}{\u}(x)$
from which we obtain the connection coefficients as
\be{21}
\omega_{\v \sp\sp \hb}^{\spp \ha}(x) = - \e{b}{\u}(x) \, \nabla_\v \,\inveT{a}{\u}(x)
= \inveT{a}{\u}(x) \, \nabla_\v \,\e{b}{\u}(x),
\ee
where the last expression follow from utilizing
$\nabla_\v\left(\e{b}{\u}(x) \, \inveT{a}{\u}(x)\right)=$ $\nabla_\v \delta^{\ha}_{\spp \hb} = 0$.
Therefore, the change in the tetrad components $\delta \inveT{a}{\u}(x)$ is computed as follows
\bea{22}
\delta \inveT{a}{\u}(x) &=& \nabla_{\mathbf{u}} \, \inveT{a}{\u}(x)\,d\tau, \no
&=& u^\v(x)\,\nabla_{\v} \, \inveT{a}{\u}(x)\,d\tau, \no
&=& -u^\v(x) \,\omega_{\v \sp\sp \hb}^{\spp \ha}(x) \, \inveT{b}{\u}(x) \, d\tau, \no
&\equiv& \chi^{\ha}_{\spp \hb}(x)\, \inveT{b}{\u}(x) \, d\tau.
\eea
where we have defined local rotation matrix
\bea{23}
\chi^{\ha}_{\spp \hb}(x) &=& -u^\v(x) \,\omega_{\v \sp\sp \hb}^{\spp \ha}(x),\no
&=& \e{b}{\u}(x)\,\nabla_{\mathbf{u}}\,\inveT{a}{\u}(x)
 =  -\inveT{a}{\u}(x)\,\nabla_{\mathbf{u}}\,\e{b}{\u}(x).
\eea

Substituting \Eq{20}, \Eq{22} and \Eq{23} into \Eq{18} yields $\delta p^{\ha}(x)$
in the the desired form of \Eq{16} with the infinitesimal LLT $\lambda^{\ha}_{\spp \hb}(x)$
given by \cite{ueda}
\be{24}
\lambda^{\ha}_{\spp \hb}(x) = \frac{1}{m} \, \left[ a^{\ha}(x) \, p_{\hb}(x)- p^{\ha}(x) \, a_{\hb}(x) \right]
                             + \chi^{\ha}_{\spp \hb}(x),
\ee
where we have used $V^{\ha}(x) = \inveT{a}{\u}(x) V^\u(x)$ for an arbitrary vector $\mbf{V}(x)$, and
$\mbf{V}(x)\cdot\mbf{V}(x) = $ $V^\u(x)\,V_\u(x)=$ $V^{\ha}(x)\,V_{\ha}(x)$.
As discussed above, the first term in \Eq{24} arises from the right hand side of the FW transport law and
involves the local rotation of the observer's 4-velocity in the $\mbf{u}(x)$-$\mbf{a}(x)$ plane only. The second term
$\chi^{\ha}_{\spp \hb}(x)$ arises when the observer chooses not to FW transport the tetrad that
defines his local reference frame. For a vector $\mbf{w}(x)$ perpendicular to the $\mbf{u}(x)$-$\mbf{a}(x)$ plane,
$\chi^{\ha}_{\spp \hb}(x)$ produces
a rotation in the plane perpendicular to $\mbf{u}(x)$ and $\mbf{w}(x)$, i.e
$\mbf{\chi}(x)\cdot\mbf{u}(x) = 0$ and $\mbf{\chi}(x)\cdot\mbf{w}(x) = 0$.
If $\mathbf{e}'_{\hi}(x)$ are the spatial axes of a non-FW transported tetrad, and $\mathbf{e}_{\hi}(x)$ are
the spatial axes of a second FW transported tetrad, then the former will precess locally relative to the
latter with an angular velocity vector equal to $\mbf{w}(x)$ (see MTW, p174 in \cite{tetrads}).

We can now calculate the components of the local Wigner transformation $W^{\ha}_{\spp \hb}(x)$ that appear
in \Eq{14} which determines how the spin of the particle rotates locally
$(\ket{p^{\hi}(x),\sigma} \to \ket{p^{\,' \hi}(x),\sigma'} = U(\Lambda(x))\ket{p^{\hi}(x),\sigma}$)
as the particle traverses from $x^\u \to x^{\, ' \u}$ along its trajectory in curved spacetime.
Performing this calculation to first order in $d\tau$ using
\be{25}
W^{\ha}_{\spp \hb}(x) \equiv \delta^{\ha}_{\spp \hb} + \vartheta^{\ha}_{\spp \hb}(x)\, d\tau,
\ee
in the definition of the Wigner rotation \Eq{15}, and additionally
\Eq{17} for the form of an arbitrary LT to $\O(d\tau)$, one can derive (after a lengthy calculation)
the following expression for the infinitesimal
local Wigner rotation $\vartheta^{\ha}_{\spp \hb}(x)$  \cite{ueda}
\bea{26}
 \vartheta^{\ho}_{\sph \ho}(x) &=& \vartheta^{\ho}_{\sph \hi}(x) = \vartheta^{\hi}_{\sph \ho}(x) = 0, \no
 \vartheta^{\hi}_{\sph \hj}(x) &=& \lambda^{\hi}_{\sp \hj}(x)
 +\frac{\lambda^{\hi}_{\sph \ho}(x)\, p_{\hj}(x) - p^{\hi}(x) \,\lambda_{\hj\,\ho}(x) }{p^{\ho}(x)+m}.
\eea
A derivation of \Eq{26} is given in Appendix~\ref{appA}. For a particle of spin-$j$ the
rotation matrix $D^{(j)}_{\sigma'\sigma}(W(x))$ that appears in \Eq{14} is given to $\O(d\tau)$ by
\be{27}
D^{(j)}_{\sigma'\sigma}(W(x)) = I + i\,\left[ \vartheta_{\hat{2}\hat{3}}(x)\,J_{\hat{1}}
+ \vartheta_{\hat{3}\hat{1}}(x)\,J_{\hat{2}} + \vartheta_{\hat{1}\hat{2}}(x)\,J_{\hat{3}} \right]\, d\tau
\ee
where
$\left[J_{\hi},  J_{\hj}\right] = i \epsilon_{\hi\hj\hk}\,J_{\hk}$ are the commutation relations for $SU(2)$
with the constant flat spacetime spin-$j$ matrices $\{J_{\hi}\}$.

For the case of spin-$\half$, to which we now specialize,
 we have $\{J_{\hi} = \half\sigma_{\hi}\}$ where $\{\sigma_{\hi}\}$ are the usual flat
spacetime constant Pauli matrices. The infinitesimal unitary transformation of the
state $\ket{p^{\hi}(x),\sigma}$ as the particle moves from $x^\u\to x^{'\u}$ is given by \cite{ueda}
\begin{subequations}
\bea{28a}
U(\Lambda(x))\,\ket{p^{\hi}(x),\uparrow} &=&
\left(I + \frac{i}{2}\,\vartheta_{\hat{2}\hat{3}}(x)\,d\tau\right) \, \ket{p^{\hi}(x'),\uparrow} \no
&-&\half\,\Big(\vartheta_{\hat{3}\hat{1}}(x) - i \vartheta_{\hat{2}\hat{3}}(x)\Big)\,d\tau \ket{p^{\hi}(x'),\downarrow}, \\
U(\Lambda(x))\,\ket{p^{\hi}(x),\downarrow} &=&
\half\,\Big(\vartheta_{\hat{3}\hat{1}}(x) + i \vartheta_{\hat{2}\hat{3}}(x)\Big)\,d\tau \ket{p^{\hi}(x'),\uparrow} \no
&+&\left(I - \frac{i}{2}\,\vartheta_{\hat{2}\hat{3}}(x)\,d\tau\right) \, \ket{p^{\hi}(x'),\downarrow},\label{28b}
\eea
\end{subequations}
where we have used the notation $\sigma = \{\half,-\half\} = \{\uparrow,\downarrow\}$.

We can iterate the formula for the infinitesimal local Wigner rotation to obtain the finite rotation
between an initial and final point in spacetime $\{x(\tau_i),x(\tau_f)\}$. Breaking up the trajectory
into $N$ infinitesimal time steps of length $\tau_{i,f}/N$ where
$\tau_{f,i} = \int_{\tau_i}^{\tau_f}\,d\tau =$ $\int_{\tau_i}^{\tau_f}\, (g_{\u\v}(x) \, dx^\u \, dx^\v)^{1/2}$
is the total proper time between the two events, and $x^\u_k = x^\u(\tau_i + k\tau_{f,i}/N)$
\bea{29}
W^{\ha}_{\spp \hb}(x_f,x_i) &=& \lim_{N\to\infty}\prod_{k=0}^{N}
\left[\delta^{\ha}_{\sph \hb} + \vartheta^{\ha}_{\sph \hb}(x_k) \frac{\tau_{f,i}}{N}\ \right], \no
&=& T \exp\left[\int_{\tau_i}^{\tau_f} d\tau \,\vartheta^{\ha}_{\sph \hb}(x(\tau))\right].
\eea
In the last expression the time ordering operator $T$ is required since, in general, the
infinitesimal local Wigner rotations $\vartheta^{\ha}_{\sph \hb}(x(\tau))$ do not commute at different
locations $x^\u(\tau)$ along the trajectory.

Some immediate observations can be made from the above formulas for the Wigner rotation matrix \Eq{26},
which depends on the infinitesimal LT matrix $\lambda^{\ha}_{\spp \hb}(x)$ \Eq{24} and the particle's 4-momentum.
If we choose the observer's local laboratory to ride along with the particle, i.e. by selecting the observer's temporal tetrad
vector to be equal to the particle's 4-velocity $\bfe{0}(x) = \mbf{u}(x)$, the particle is observed to be
instantaneously at rest. The local spatial components of the particle's 4-momentum are then zero since
$$
p^{\ha}(x) = p^\u(x)\,\inve{\u}{a}(x)\, = m \,\e{0}{\u}(x)\,\inve{\u}{a}(x) = m \,\delta_{\ho}^{\spp \ha}, \quad
\Rightarrow \quad p^{\hi}(x) = 0.
$$
%
Since the non-trivial infinitesimal portion of the Wigner rotation matrix $\vartheta^{\hi}_{\,\hj}(x)$ \Eq{26}, depends only
upon the spatial components of the local 4-momentum  $p^{\hi}(x)$, the former reduces to
$\vartheta^{\hi}_{\,\hj}(x) = \chi^{\hi}_{\,\hj}(x)$
(since, for this case, $\lambda^{\hi}_{\,\hj}(x) = \chi^{\hi}_{\,\hj}(x)$ from \Eq{24}, for the same reason).
From \Eq{23} we have
$\chi^{\ha}_{\spp\hb}(x)=-\inveT{a}{\u}(x)\,\nabla_{\mathbf{u}}\,\e{b}{\u}(x)$, which vanishes, in particular, for the case
of geodesic motion in which the 4-acceleration $\mbf{a}(x)=0$, and the choice of a FFF tetrad \Eq{9}. For this
tetrad choice the Wigner rotation matrix \Eq{25} reduces to the identity matrix
$W^{\ha}_{\spp \hb}(x) = \delta^{\ha}_{\spp \hb}$, i.e. from this frame the observer detects no Wigner rotation
of the particle's spin. Note that the non-rotation of the particle's spin depended upon the particular choice of
the observer's reference frame, i.e. the FFF tetrad for geodesic motion discussed above. Any other choice
of the spatial tetrad vectors would yield a non-trivial Wigner rotation
$\vartheta^{\hi}_{\,\hj}(x) = \chi^{\hi}_{\,\hj}(x)$ for the geodesic motion, in which $\chi^{\hi}_{\,\hj}(x)$
describe the spatial rotations of the observer's reference frame relative to the FFF.

For the case of arbitrary motion the particle, with tangent $\mbf{u}$ and
$\mbf{a} = \nabla_{\mathbf{u}}\,\mbf{u}\ne 0$, does not move on a geodesic.
However, in analogy to the above discussion, a similar reference frame can be found for which the observer detects no
Wigner rotation. The observer again rides along in the instantaneous rest frame of the particle, with the selection of
$\bfe{0}(x) = \mbf{u}(x)$ implying that $p^{\hi}(x)=0$ and $\vartheta^{\hi}_{\,\hj}(x) = \chi^{\hi}_{\,\hj}(x)$.
The specific choice of tetrad is governed by the requirement that it is FW transported along the particle's
trajectory. This leads to the vanishing $\chi^{\hi}_{\,\hj}(x)$, the space-space portion of $\chi^{\ha}_{\spp \hb}(x)$.
This can be seen a follows: let $\mbf{V}$ in the FW transport equation \Eq{10} be any of the four
tetrad vectors $\bfe{b}(x)$. In component form, \Eq{10} becomes
$$
\nabla_{\mathbf{u}} \,\e{b}{\u}(x) = [\e{b}{\v}(x)\,u_\v(x)]\,a^\u(x) - [\e{b}{\v}(x)\,a_\v(x)]\,u^\u(x).
$$
Multiplying this expression by $-\inve{\u}{a}(x)$ and using the definition
$\chi^{\ha}_{\spp \hb}(x) = -\inve{\u}{a}(x)\,\nabla_{\mathbf{u}} \,\e{b}{\u}(x)$
in \Eq{23} yields
$$
\chi^{\ha}_{\spp \hb}(x) = - [\e{b}{\v}(x)\,u_\v(x)]\,[a^\u(x)\,\inve{\u}{a}(x)]
                        + [\e{b}{\v}(x)\,a_\v(x)]\,[u^\u(x)\,\inve{\u}{a}(x)].
$$
Finally, substituting the expression for the choice of the temporal tetrad as the 4-velocity of the particle for
the instantaneous rest frame of the particle, $u^\u(x) = \e{0}{\u}(x)$ and $u_\v(x) = \inve{\v}{0}(x)$, and
using the orthonormality of the tetrad vectors \Eq{3} produces the expression
\bea{30}
\chi^{\ha}_{\spp \hb}(x) &=& - \delta_{\hb}^{\spp \ho}\,[a^\u(x)\,\inve{\u}{a}(x)]
                           + [\inve{b}{\v}(x)\,a^\v(x)]\,\delta_{\ho}^{\spp \ha} \no
& = & - a^{\ha}(x) \, \delta_{\hb}^{\spp \ho} + \delta_{\ho}^{\spp \ha}\,a_{\hb}(x),
 \quad \trm{in the FWF}
\eea
which implies $\chi^{\hi}_{\,\hj}(x)=0$ (the non-zero time-space components
$\chi^{\ho}_{\spp \hi}(x) = -\chi^{\hi}_{\spp \ho}(x)$ $= a_{\hi}(x)$
describe local spatial acceleration (boost) in the $\mbf{u}$-$\mbf{a}$ plane), and hence a null Wigner rotation in this
FW transported reference frame (FWF for short). Any other choice of tetrad would lead to the observer detecting a non-trivial Wigner rotation, relative
to the FWF.

In the following sections we will investigate the $\O(\hbar)$ correction to
the particle's 4-velocity and 4-acceleration
when the classical geodesic motion of a spin-$\half$ particle (defined by prescribing $\mbf{u}$
independent of the particle's spin such that $\nabla_{\mathbf{u}}\,\mbf{u}=0$),
is replaced by the quantum mechanical Dirac current of the particle. To accomplish this task, we must
first discuss  the Dirac equation (DE) in curved spacetime \cite{DE_in_GR}, which we turn to in the next section.

\section{The Dirac Equation in curved spacetime}
The covariant derivative $\nabla_\a$ discussed in the previous section transforms world tensors into world tensors
under GCTs. The tetrad and inverse tetrad components allow us to relate the components of a world tensor $T^{\a}_{\spp \b \c}(x)$
to the observer's LIF components $T^{\ha}_{\spp \hb\hc}(x)$ by \Eq{21}. Therefore, it would be desirable to introduce
the concept of a derivative that is covariant (i.e. transforms well) under LLTs.  For world tensors, the affine connection \Eq{7}
was introduced on the spacetime to define the notion of parallel transport. For a world vector $V^\a(x)$ at a spacetime
point $P$ with coordinates $\mbf{x}$, one defines the \tit{parallel translate} $V_{\parallel}^\a(x\to x+dx)$ at point $Q$ with
coordinates $\mbf{x}+d\mbf{x}$ of the vector $V^\a(x)$ at $P$ by
$V_{\parallel}^\a(x\to x+dx) = V^\a(x) - \Gamma^{\a}_{\spp \l\b}(x) \, V^\l(x)\,dx^\b$. The covariant derivative $\nabla_\b V^\a(x)$
is then defined as the subtraction of the vector ``already at the point $Q$", $V^\a(x+dx)$, and the parallel translate
at $Q$, $V_{\parallel}^\a(x\to x+dx)$, in the limit that $Q$ approaches $P$
\be{31}
\nabla_\b V^\a(x) = \lim_{dx\to 0} \frac{V^\a(x+dx) - V_{\parallel}^\a(x\to x+dx)}{dx}
=\pd{\b} V^\a(x) + \Gamma^{\a}_{\spp \l\b}(x) \, V^\l(x).
\ee

We now wish extend above concept of a covariant derivative for world vectors to
local Lorentz vectors $V^{\ha}(x)$. We do this by introducing a local \tit{spin connection}
$\spinc{\u}{\ha}{\hb}(x)$ used to define the parallel translate $V_{\parallel}^{\ha}(x\to x+dx)$
at the point $Q$, of a local Lorentz vector $V^{\ha}(x)$ at point $P$, by
$V_{\parallel}^{\ha}(x\to x+dx) = V^{\ha}(x) -  \spinc{\u}{\ha}{\hb}(x)\, V^{\hb}(x)\,dx^\u$.
In the analogy with world vectors, the \tit{local} covariant derivative $\D_\u V^{\ha}(x)$is defined by
\be{32}
\D_\u V^{\ha}(x) = \lim_{dx\to 0} \frac{V^{\ha}(x+dx) - V_{\parallel}^{\ha}(x\to x+dx)}{dx}
=\pd{\u} V^\a(x) + \spinc{\u}{\ha}{\hb}(x)\, V^{\hb}(x).
\ee
One could further introduce the notation $\D_{\ha} \equiv \e{a}{\u}(x)\,\D_{\u}$,
$\pd{\ha} \equiv \e{a}{\u}(x)\,\pd{\u}$ and $\spinc{\hc}{\ha}{\hb}(x) = \e{c}{\u}(x)\,\spinc{\u}{\ha}{\hb}(x)$
so that \Eq{32} only contains LIF (Latin) indices. However, such notation connotes the existence of
a set of local coordinates $y^{\ha}(x^\u)$. Such coordinates do exist, centered on the
origin of the observer's laboratory (e.g. Riemann normal coordinates), but
as discussed in the previous section, have only a limit range of applicability. We will retain
the notation $\D_\u$ to emphasize our interest in the change of local Lorentz quantities as we move
from point to point $x^\a\to x^\a + dx^\a$ in the curved spacetime.

By requiring that \Eq{32} is ``compatible" with \Eq{31} in the sense that we
can transform between equations using the tetrad/inverse tetrad \Eq{4}, we obtain
the equation for the spin connection given by \Eq{21} (see Lawrie \cite{lawrie}).
Equation~(\ref{21}) can be rearranged into the following form
\be{33}
D_\u \e{a}{\v}(x) \equiv \pd{\u} \e{a}{\v}(x)
+ \Gamma^{\v}_{\spp \l\u}(x)\, \e{a}{\l}(x) - \spinc{\u}{\hb}{\ha}(x) \, \e{b}{\v}(x)=0,
\ee
which defines the \tit{total covariant derivative} which transforms properly under
both GCTs  and LLTs, with a Christoffel connection for every Greek index and
a spin connection for every Latin index, respectively (note: the covariant
derivative $\nabla_\b$ in \Eq{31} acts only on world (Greek) indices, while the
local covariant derivative $\D_\u$ in \Eq{32} acts only on local (Latin) indices).
Equation~(\ref{33}) can be considered a compatibility requirement of the tetrad
(the \tit{first veirbein postulate}, see Ortin, \cite{ortin}) analogous to the
metric compatibility equation $\nabla_\u g_{\a\b}(x)=0$ which defined the world Christoffel connection.

In general, the spin connection can be defined completely in terms of the (orthonormal) tetrad
defining the observer's LIF, via
\be{34}
\omega_{\ha\hb\hc}(x) = - \Omega_{\ha\hb\hc}(x) +\Omega_{\hb\hc\ha}(x) + \Omega_{\ha\hb\hc}(x),
\ee
where
\be{35}
[ \bfe{a}(x), \bfe{b}(x) ] = -2 \Omega_{\ha\hb}^{\sp\sp\sp \hc}(x)\, \bfe{c}(x), \quad
\Omega_{\ha\hb}^{\sp\sp\sp \hc}(x) = \e{a}{\u}(x)\,\e{b}{\v} \, \pd{\,[\u}\inveT{c}{\v]}(x)
\ee
define the \tit{Ricci rotation coefficients} which are the commutators of the basis vectors
$\bfe{a}(x) = \e{a}{\u}(x)\,\pd{\u}$. A \tit{non-holonomic} frame is one in which the $\Omega$s
do not vanish, while a coordinate basis is one in which they do (i.e. $\e{a}{\u}(x) = \delta_{\ha}^{\spp \u}$).

So far the above discussion has been in terms of vectors and tensors. The utility of the tetrad formalism is
that it allows one to introduce spinor (integer and half-integer) representations of the Lorentz group through
the spin connection (so named), which is necessary in order to describe fermions. This is the only known method
by which to describe spinors in curved spacetime in arbitrary coordinates, and thus the only known method
to couple fermions to gravity (see Ortin, \cite{ortin}). If we denote a general spinorial quantity in the LIF of
spin-$j$ by $\Psi^A(x)$ with $A=(-j,-j+1,\ldots,j)$ taking on $2j+1$ values, then total covariant derivative
of $\Psi^A(x)$ is given by
\be{36}
D_\u \,\Psi^A(x) = \pd{\u} \Psi^A(x) - \half \spincup{\u}{\hb}{\hc}(x) \,\Gamma_{(j)}(\Sigma_{\hb\,\hc})^A_{\spp B}\,\Psi^B(x).
\ee
In \Eq{36} $\Gamma_{(j)}(\Sigma_{\hb\,\hc})$ is the matrix representation for spin-$j$ of the flat spacetime
generators $\Sigma_{\hb\,\hc}$ of the Lorentz group. These are constant matrices that satisfy the commutation
relations
\be{37}
[\Sigma_{\ha\,\hb}, \Sigma_{\hc\,\hd}] = - \eta_{\ha\hc}\,\Sigma_{\hb\hd} - \eta_{\hb\hd}\,\Sigma_{\ha\hc}
                                         + \eta_{\ha\hd}\,\Sigma_{\hb\hc} + \eta_{\hb\hc}\,\Sigma_{\ha\hd}.
\ee
For vectors we have the representation
$\Gamma_{(1)}(\Sigma_{\hc\,\hd})^{\ha}_{\spp \hb} = 2 \eta_{\,[\hc}^{\spp \ha} \, \eta_{\hd]\,\hb}$, while
for spinors ($j=1/2$) we have
$\Gamma_{(1/2)}(\Sigma_{\ha\,\hb}) =$ $\half\gamma_{[\ha},\gamma_{\hb]}=$ $\frac{1}{4} \left[\gamma_{\ha},\gamma_{\hb}\right]$,
where $\gamma_{\ha} = \eta_{\ha\hb}\, \gamma^{\hb}$ are the usual flat spacetime Dirac gamma matrices
satisfying the anti-commutation relations $\{\gamma_{\ha},\gamma_{\hb}\} = 2 \eta_{\ha\hb}$.
These representations of the Lorentz group lead to the following formulas for the total covariant
derivative of LIF vectors and spinors, respectively \cite{ortin}
\bea{38}
D_\mu \,V^{\ha}(x) &=& \pd{\u} \, V^{\ha}(x)
- \half \spincup{\u}{\hb}{\hc}(x) \,\Gamma_{(1)}(\Sigma_{\hb\,\hc})^{\ha}_{\spp \hd}\,V^{\hd}(x), \no
&=& \pd{\u} \, V^{\ha}(x) + \spinc{\u}{\ha}{\hb}(x) \, V^{\hb}(x),
\eea
and
\bea{39}
D_\u \,\psi^A(x) &=& \pd{\u} \psi^A(x)
- \half \spincup{\u}{\hb}{\hc}(x) \,\Gamma_{(1/2)}(\Sigma_{\hb\,\hc})^A_{\spp B}\,\psi^B(x), \no
&=& \pd{\u} \, \psi^{A}(x) -\frac{1}{8} \spincdn{\u}{\ha}{\hb}(x)\,([\gamma^{\ha},\gamma^{\hb}])^{A}_{\spp B}  \, \psi^{B}(x),\no
&\equiv& \pd{\u} \, \psi^{A}(x) + (\Gamma_\mu)^{A}_{\spp B}(x)  \, \psi^{B}(x),
\eea
where we have defined the \tit{spinor connection} $\Gamma_\u(x)$ for $j=1/2$
\be{40}
\Gamma_\mu(x) = -\frac{1}{8} \spincdn{\u}{\ha}{\hb}(x)\,[\gamma^{\ha},\gamma^{\hb}]
= \frac{i}{4} \spincdn{\u}{\ha}{\hb}(x)\,\sigma^{\ha\hb}, \quad
\spincdn{\u}{\ha}{\hb}(x) = \inve{\u}{c}(x)\,\spincdn{\hc}{\ha}{\hb}(x),
\ee
using the conventional definition of $\sigma^{\ha\hb} = \frac{i}{2}\,[\gamma^{\ha},\gamma^{\hb}]$,
and transforming the first index of $\spincdn{\hc}{\ha}{\hb}(x)$ from \Eq{34}
using the inverse tetrad components.
Henceforth, we shall suppress the explicit spinor indices (unless needed) and write \Eq{39} as
$D_\u \,\psi(x) = (\pd{\u} + \Gamma_\u)\,\psi(x)$. The promotion of the Dirac equation in flat spacetime
(with global coordinates $x^{\ha}$)
$$
\left(i \,\gamma^{\ha}\,\pd{\ha} - m/\hbar\right)\, \psi(x) = 0, \quad \trm{flat spacetime}
$$
to curved spacetime, using the \tit{minimal coupling} prescription $\partial_{\ha}\to D_{\ha}$, becomes
\bea{41}
\lefteqn{
\left(i \,\gamma^{\ha}\,D_{\ha} - m/\hbar\right)\, \psi(x)
= \left(i \,\gamma^{\a}(x)\,D_{\a} - m/\hbar\right)\, \psi(x)} \no
&=& \big(i \,\gamma^{\a}(x)\,[\pd{\a} + \Gamma_\a(x)] - m/\hbar\big)\, \psi(x) = 0, \quad \trm{curved spacetime}
\eea
in world coordinates $x^{\a}$, where we have defined the curved spacetime Dirac gamma matrices by
\be{42}
\gamma^{\a}(x) = \gamma^{\ha} \, \e{a}{\a}(x),
\ee
and we have used $\gamma^{\ha}\,D_{\ha} =$ $ \gamma^{\ha}\,\e{a}{\a}(x)\,D_{\a} =$ $\gamma^{\a}(x)\,D_{\a}$.

The FFF observer in curved spacetime can always construct coordinates such that the metric $g_{\a\b}(x)$ and
the Christoffel symbols $\Gamma^{\u}_{\spp \a\b}(x)$ vanish along the geodesic trajectory
(e.g. Riemann normal coordinates, FW normal coordinates, \ldots). However, the observer cannot in general
choose coordinates so that all the second derivatives of the metric vanish along the geodesic, unless the
spacetime is flat. This is described by the Riemann curvature tensor, which has its most direct definition
in terms of the commutator of the covariant derivative. For vectors and spinors in the LIF this can be
defined as
\be{43}
[\, D_\u, D_\v\,] V^{\ha}(x) = R_{\u\v\hb}^{\spp\spp\sp \ha}(\omega(x))\, V^{\hb}(x),
\ee
and
\be{44}
[\, D_\u, D_\v\,] \psi(x) = \frac{i}{4}\,R_{\u\v\ha\hb}(\omega(x))\,\sigma^{\ha\hb}\, \psi(x)
= \frac{i}{4}\,R_{\u\v\c\d}(\Gamma(x))\,\sigma^{\c\d}(x)\, \psi(x).
\ee
In the above, $R_{\u\v\hb}^{\spp\spp\sp \ha}(\omega)$ is the Riemann curvature tensor formed from
the spin connection, and is related to the usual Riemann curvature world tensor $R_{\u\v\a}^{\spp\spp\sp \b}(\Gamma)$
constructed from the Christoffel symbols by
\be{45}
R_{\u\v\a}^{\spp\spp\sp \b}(\Gamma(x)) = \inveT{a}{\a}(x)\,\e{b}{\b}(x)\,R_{\u\v\ha}^{\spp\spp\sp \hb}(\omega(x)),
\ee
where
\be{46}
R_{\u\v\a}^{\spp\spp\sp \b}(\Gamma(x)) =
  \pd{\,\u}\,\Gamma^{\b}_{\spp \v\a}(x)
- \pd{\,\v}\,\Gamma^{\b}_{\spp \u\a}(x)
+ \Gamma^{\b}_{\spp \u\l}(x)\,\Gamma^{\l}_{\v\a}(x)
- \Gamma^{\b}_{\spp \v\l}(x)\,\Gamma^{\l}_{\u\a}(x),
\ee
and
\be{47}
R_{\u\v\ha}^{\spp\spp\sp \hb}(\omega(x)) =
  \pd{\,\u}\,\omega_{\v\ha}^{\spp\spp \hb}(x)
- \pd{\,\v}\,\omega_{\u\ha}^{\spp\spp \hb}(x)
- \omega_{\u\ha}^{\spp\spp \hc}(x)\,\omega_{\v\hc}^{\spp\spp \hb}(x)
+ \omega_{\v\ha}^{\spp\spp \hc}(x)\,\omega_{\u\hc}^{\spp\spp \hb}(x).
\ee
Equation~(\ref{44}) and \Eq{47} will prove useful in the next section.

\section{WKB solution to the Dirac equation in curved spacetime}
\subsection{Wavefunction}
In this section we develop a WKB solution to the Dirac equation in curved spacetime \Eq{41} \cite{audretsch,alsing}.
We make no approximation to the strength or form of the gravitational field (metric), but instead only keep the
lowest $\O(\hbar)$ quantum correction to quantities of physical interest. We develop the WKB expansion of $\psi(x)$ as
\be{48}
\psi(x) = e^{i S(x)/\hbar}\,\sum_{n=0}^{\infty}\,(-i\hbar)^n\,\psi_n(x),
\ee
where the action $S(x)$ is real. Inserting this expansion into \Eq{41}, and using result that
for functions $D_\a S(x) = \pd{\a}\,S(x)$, we have, upon equating like powers of $\hbar$
\be{49}
[\gamma^\a(x)\,\pd{\a}S(x) + m]\,\psi_0(x) = 0,
\ee
\be{50}
[\gamma^\a(x)\,\pd{\a}S(x) + m]\,\psi_1(x) = -\gamma^\a(x)\,D_\a\,\psi_0(x).
\ee
Since the term in the square brackets in \Eq{49} is a matrix, the condition that this equation
has non-trivial solutions requires its determinant to be identically zero
\be{51}
\trm{det}[\gamma^\a(x)\,\pd{\a}S(x) + m] = 0.
\ee
Equation~(\ref{51}) reduces to the classical Hamilton-Jacobi equation upon using \Eq{49} twice
\bea{52}
m^2  &=& \gamma^\a(x)\,\pd{\a}\,S(x)\,\gamma^\b(x)\,\pd{\b}S(x)\,
= \half\{\gamma^\a(x),\gamma^\b(x)\}\,\pd{\a}S(x)\pd{\b}S(x) = g^{\a\b}(x)\,\pd{\a}S(x)\pd{\b}S(x)\no
&=& \partial^{\a}S(x)\pd{\a}S(x).
\eea
Defining the particle's world 4-momentum $p_\a(x)=m\,u_\a(x)$ as the normal to the surface of constant action $S(x)$ we have
\be{53}
p_\a(x)= -\pd{\a}S(x), \quad p^\a(x)\,p_\a(x) = m^2,
\ee
the later equation of which simply states the normalization of the particle's world 4-velocity
\be{54}
u^\a(x)\,u_\a(x)=1.
\ee
Thus, to $\O(1)$ in $\hbar$, the phase of any quantum mechanical particle in curved spacetime,
regardless of its spin, is given by the classical result
\be{55}
S(x) = \int p_\a(x)\,dx^\a.
\ee
The above form of the action was suggested by Stodolsky \cite{stodolsky,alsing} who pointed out
that the action of a free particle is given by $S(x) = m\int ds$. Writing the line element as
$ds = g_{\a\b}(x) dx^\a dx^\b/ds$ and defining $p_\a(x) = m g_{\a\b}(x) dx^\b/ds = m u_\a(x)$
reproduces \Eq{55}.

The determinant condition \Eq{51} and the resulting Hamilton-Jacobi equation \Eq{52} arise
in the solution of the DE in flat spacetime. As such, in the observer's local FFF the
general solution to \Eq{49} takes the flat spacetime form
\be{56}
\psi_0(x) = \beta_{\uparrow}(x)\,  \psi^{(\uparrow)}_0(x) + \beta_{\downarrow}(x)\,  \psi^{(\downarrow)}_0(x)
\ee
where $\beta_{\uparrow}(x)$ and $\beta_{\downarrow}(x)$ are scalar functions and
the positive energy spin up $ \psi^{(\uparrow)}_0(x)$ and spin down $ \psi^{(\downarrow)}_0(x)$
Dirac spinors \cite{mandl_shaw} are given by
\be{57}
 \psi^{(\uparrow)}_0(x) = \left(\frac{E+m}{2m}\right)^{1/2} \,
\left(
\begin{array}{c}
  1 \\
  0 \\
  \displ\frac{p^{\hat{3}}}{E+m} \\
  \displ\frac{p^{\hat{1}} + i p^{\hat{2}}}{E+m} \\
\end{array}
\right),
\quad
\psi^{(\downarrow)}_0(x) = \left(\frac{E+m}{2m}\right)^{1/2} \,
\left(
\begin{array}{c}
  0 \\
  1 \\
  \displ\frac{p^{\hat{1}} - i p^{\hat{2}}}{E+m} \\
  \displ-\frac{p^{\hat{3}}}{E+m} \\
\end{array}
\right)
\ee
with
\be{58}
\big( E(x), \,p^{\hi}(x)\big) \equiv p^{\ha}(x) = p^\a(x) \, \inve{\a}{a}(x).
\ee
Equation~(\ref{58}) states that, in general (i.e. not necessarily a FFF),
an observer carrying local tetrad axes $\bfe{a}$
measures the world 4-momentum $p^\a(x)$ of a particle (massive or massless) crossing his laboratory
by projecting $p^\a(x)$ onto his local axes (in the above using components of the inverse tetrad).
By \Eq{1} and \Eq{2}, the observer's axes form an orthonormal basis, so that the metric is
locally flat $\eta_{\ha\hb}$ for the observer. Thus, the first equality in \Eq{58} states that the
components of $p^\a(x)$ measured by the observer take the usual special relativistic flat spacetime form
$p^{\ho}=E = m/(1-v^2(x))$ and $p^{\hi} = m v^{\hi}(x)/(1-v^2(x))$ were $v^{\hi}(x)$ are the spatial components
of the locally measured velocity.

For the particular case of the freely falling frame
that we are considering, $p^\a(x)$ is the world 4-momentum of the observer's origin, where the spinor resides,
This is the curved spacetime generalization of the
particle's rest frame. As such, by the orthonormality of tetrad \Eq{3} we have
\be{59}
p^{\ha}(x) = \big( E(x), \,p^{\hi}(x)\big) = (m,0,0,0), \quad \trm{in a FFF}.
\ee
As the observer traverses his geodesic trajectory in the curved spacetime, the components of his 4-momentum
at each spacetime point $x^\a$ will have the form of \Eq{59} in the FFF.
Thus, in the FFF the normalized, positive energy Dirac spinors of \Eq{57} take the rest-frame form
\be{60}
 \psi^{(\uparrow)}_0(x) =
\left(
\begin{array}{c}
  1 \\
  0 \\
  0 \\
  0 \\
\end{array}
\right),
\qquad
\psi^{(\downarrow)}_0(x) =
\left(
\begin{array}{c}
  0 \\
  1 \\
  0 \\
  0 \\
\end{array}
\right)
\ee

It can be shown \cite{audretsch} that the general zeroth order solution to the Dirac equation in curved spacetime $\psi_0(x)$
can be written as an amplitude times a normalized spinor $\psi_0(x) = f(x) \,\varphi_0(x)$ (see also \cite{alsing}),
where $f(x)$ satisfies the equation $\pd{a} f(x) \, u^\a(x) = -\theta(x) \,f(x)$, and $\theta(x) \equiv \nabla_\a u^\a(x)$ is
the expansion of the cross section of a congruence (``tube") of timelike geodesics \cite{optical_scalars}. The
normalized spinor $\varphi_0(x)$
(i.e. $\bar{\varphi}_0(x)\,\varphi_0(x) =1$ where $\bar{\varphi}_0(x) = \varphi^\dagger(x) \gamma^{\ho}$), satisfies
$u^\a(x) D_\a \varphi_0(x)=0$ which states that $\varphi_0(x)$ is parallel propagated along the congruence.
The results are most easily proved by introducing Riemann normal coordinates along an arbitrarily chosen timelike
geodesic of the congruence along which the spin-$\half$ travels and for which the spinor connection $\Gamma_\a(x)$ vanishes
(see \Eq{40} and \Eq{41}).

In this work, we will not introduce such local laboratory coordinates (which have limited spacetime ranges) and instead
perform computations in the world coordinates $x^\a$ of the metric $g_{\a\b}(x)$, and project world tensors to LIF tensors
by means of the tetrad, as in \Eq{5}. It is easy, and instructive to show that $\varphi_0(x)$ is parallel transported
along the congruence in general coordinates. Taking $\varphi_0(x) = \psi^{(\uparrow)}_0(x)$ in the FFF from \Eq{60},
with no loss in generality, we have
\Bea
u^\a(x) D_\a \psi^{(\uparrow)}_0(x) &=& u^\a(x)[\pd{\a} + \Gamma_\a(x) ]\,\psi^{(\uparrow)}_0(x)
                                     = u^\a(x) \Gamma_\a(x) \,\psi^{(\uparrow)}_0(x),\\
&=&\frac{i}{4} u^\a(x) \omega_{\a\ha\hb}(x) \sigma^{\ha\hb}\,\psi^{(\uparrow)}_0(x)
=\frac{i}{4} \eta_{\ha\hc} \inveT{c}{\u}(x) \big[ u^\a(x) \nabla_\a \e{b}{u} \big] \sigma^{\ha\hb}\,\psi^{(\uparrow)}_0(x) = 0,
\Eea
where in the second and third equalities we have used the constancy of $\psi^{(\uparrow)}_0(x)$ in the FFF,
the definition of the spinor connection $\Gamma_\a(x)$ in \Eq{40} and the definition of the
spin connection $ \omega_{\a\ha\hb}(x)$ in \Eq{21}. The last equality follows from the definition of
the FFF in \Eq{9}, that the entire tetrad is parallel transported along the worldline of the FFF.
Further, we note that from \Eq{23},  $\chi^{\ha}_{\spp \hb}(x)=0$ in the FFF and additionally,
$u^\a(x) \Gamma_\a(x) =  u^{\ha}(x) \Gamma_{\ha}(x) = \e{0}{\ha}(x)\Gamma_{\ha}(x) = \Gamma_{\ho}(x)=0$,
while the spatial $\Gamma_{\hi}(x)$ are in general non-zero, unless one uses locally adapted coordinates
(e.g. Riemann normal coordinates).

For the $\O(\hbar)$ solution to the DE, one notes that the operator in
the square brackets in the non-homogeneous \Eq{50} is the
same one that appears in the $\O(1)$ homogeneous \Eq{49}. In order for \Eq{50} to have a non-trivial solution
$\psi_1(x)$, all the solutions of the corresponding transposed homogeneous equation \Eq{49} must be
orthogonal to the inhomogeneity on the right hand side of \Eq{50} (Fredholm alternative). This solvability condition for \Eq{50}
becomes \cite{audretsch} $\psi^{(\sigma)}_0(x) \gamma^\a(x) D_\a \psi_0(x)=0$, for $\sigma = \{\uparrow,\downarrow\}$.
In the following, we will only need the $\O(1)$ solutions of \Eq{49} for calculating the $\O(\hbar)$ corrections
the 4-velocity and 4-acceleration of the spin-$\half$ particle's trajectory. We therefore take as our
solution to the DE in curved spacetime the wavefunction
\be{61}
\psi(x) = \psi^{(\sigma)}_0(x) \, \exp\left[\displ\frac{i}{\hbar} \int p_\u(x) dx^\u\right], \quad \sigma = \{\uparrow,\downarrow\},
\qquad \bar{\psi}(x) \psi(x) = 1.
\ee
with $\psi^{(\sigma)}_0(x)$ given by \Eq{60} in the FFF.

\subsection{Quantum corrections to the classical 4-velocity and 4-acceleration}
\subsubsection{4-velocity}
The designation of $u^\a(x)$ as the particle's 4-velocity arises from the equivalence principle for
classical particles without spin. To exhibit the influence of the particle's quantum mechanical
spin on its trajectory one postulates that the motion of the spin-$\half$ particle through the
curved spacetime is determined by its conserved Dirac probability current $j^\a(x)$ \cite{audretsch}
\be{62}
j^\a(x) = \bar{\psi}(x) \gamma^\a(x) \psi(x), \qquad D_\a j^\a(x) = 0.
\ee
Performing a Gordon decomposition \cite{baym}, the current can be written as the sum
\be{63}
j^\a(x) = j^\a_c(x) + j^\a_M(x)
\ee
of a convection current $j^\a_c(x)$, and an internal magnetization current $j^\a_M(x)$
defined by the minimal coupling curved spacetime generalization of their flat spacetime definitions
\be{64}
j^\a_c(x) = \frac{\hbar}{2 m i} \, \left[ \bar{\psi}(x) D_\a \psi(x) - \big(D_\a \bar{\psi}(x)\big) \, \psi(x) \right],
\ee
and
\be{65}
j^\a_M(x) = \frac{\hbar}{2m} \, \bar{\psi}(x)\,\sigma^{\a\b}(x) \,\psi(x), \qquad \sigma^{\a\b}(x) = \frac{i}{2}\,[\gamma^\a(x),\gamma^\b(x) ].
\ee

\begin{figure}[h]
\includegraphics[width=3.75in,height=2.75in]{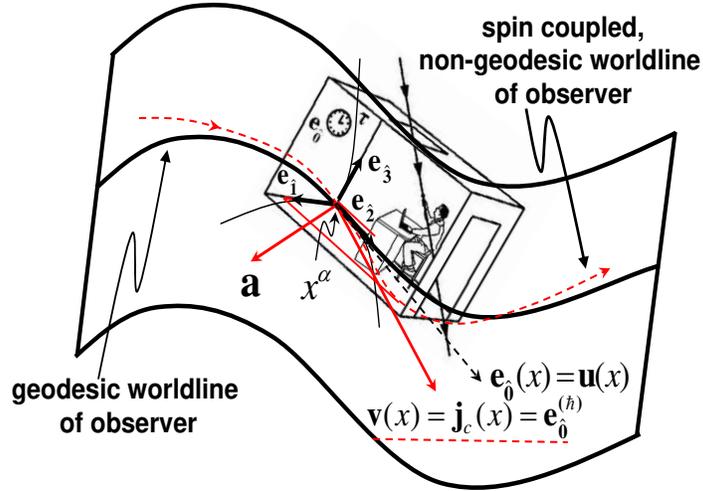}
\caption{(Color online) The middle solid line represents the geodesic trajectory ($\mathbf{a}(x)=0$) of the observer, riding along with the particle,
with 4-velocity $\bfe{0}(x)=\mathbf{u}(x)$ (black dashed arrow), if the spin of the particle is ignored
(or considered as spin 0). The dashed red line represents the perturbed, non-geodesic trajectory ($\mathbf{a}(x)\ne 0$)
with tangent $\mathbf{v}(x)$ (solid red arrow), when the spin of the particle, which couples to the spacetime,
is taken into account. The 4-velocity $\mathbf{v}(x)$
determines the particle's Dirac (convective) current $\mathbf{j}_c(x) = \mathbf{v}(x)=\bfe{0}^{(\hbar)}(x)$ which
quantum mechanically defines the particle's motion through the spacetime. (Only the
temporal axis of the $\O(\hbar)$-corrected FWF tetrad $\bfe{a}^{(\hbar)}(x)$, from which a null Wigner rotation is measured, is shown).
}
\label{fig2}
\end{figure}

We now let the convection current $j^\a_c(x)$ define a congruence of timelike curves for the free motion
of spin-$\half$ particles with tangent $v^\a(x)$
\be{66}
j_c^\a(x) = v^\a(x) = u^\a(x) + \O(\hbar), \qquad v^\a(x) v_\a(x) = 1 + \O(\hbar^2).
\ee
The normalization of the $\O(\hbar)$ corrected  $v^\a(x)$ (right hand equality in \Eq{66}) will be
required for a self-consistent definition of a 4-velocity, and will be explicitly demonstrated in the
examples considered in the subsequent sections.
Using \Eq{61} and $D_\a \psi(x) = \left[D_\a\psi^{(\sigma)}_0(x) + (i m/\hbar)\, u^\a(x)\,\psi^{(\sigma)}_0(x)\right] \exp(i S(x) /\hbar)$ we obtain
\be{67}
v_\a(x) = u_\a(x) + \frac{\hbar}{2 m i} \,\left[ \bar{\psi}^{(\sigma)}_0(x) D_\a \psi^{(\sigma)}_0(x)
                                             - \big(D_\a \bar{\psi}^{(\sigma)}_0(x)\big) \, \psi^{(\sigma)}_0(x)    \right]
                                             + \O(\hbar^2).
\ee
We can simplify the above formula for $v_\a(x)$ by noting that
$D_\a \psi^{(\sigma)}_0(x) = \big(\pd{\a} + \Gamma_\a(x)\big)\,\psi^{(\sigma)}_0(x) = \Gamma_\a(x)\,\psi^{(\sigma)}_0(x)$
and its adjoint $D_\a \bar{\psi}^{(\sigma)}_0(x) = -\bar{\psi}^{(\sigma)}_0(x)\Gamma_\a(x)$ where we have used
the constancy of $\psi^{(\sigma)}_0(x)$ in the FFF, $\bar{\psi}^{(\sigma)}_0(x) = \psi^{(\sigma)\dagger}_0(x) \gamma^{\ho}$
and $\gamma^{\ho}\Gamma^\dagger_\a(x) \gamma^{\ho} = -\Gamma_\a(x)$, which yields
\be{68}
v_\a(x) = u_\a(x) + \frac{\hbar}{m i} \; \bar{\psi}^{(\sigma)}_0(x) \, \Gamma_\a(x) \, \psi^{(\sigma)}_0(x) + \O(\hbar^2).
\ee
\Fig{fig2} illustrates the spin-coupled non-geodesic trajectory of the particle with 4-velocity $\mathbf{v}(x)$
perturbed away from its geodesic trajectory with 4-velocity $\mathbf{u}(x)$ when the spin of the particle is ignored.
\subsubsection{4-acceleration}
We next compute the quantum mechanically corrected acceleration from the velocity correction \Eq{67}
\be{69}
a_\a(x) = v^\b(x)\,D_\b v_\a(x) = 2 v^\b(x)\,D_{[\b} \,v_{\a]}(x),
\ee
which is the generalization of \Eq{8} when the 4-velocity contains both world vectors and spinors. The second
equality in \Eq{69} follows from differentiating the normalization condition for $v^\a(x)$ which allows
one to add to the middle expression an identically zero term of the form $v^\b(x)D_\a v_\b(x)=0$. In
differentiating $v_\a(x)$, one must be mindful of the way the total covariant derivative $D_\b$ acts
on the individual vector \Eq{38} and spinor terms \Eq{39}. A detailed derivation given in Appendix~\ref{appB}
reveals that the only nonzero terms that remain upon differentiation of $v_{\a}(x)$ are
\bea{70}
a_\a(x) &=& 2 v^\b(x)\,\left[ \bar{\psi}^{(\sigma)}_0(x) D_{[\b}, D_{\a]} \psi^{(\sigma)}_0(x)
         - \big(D_{[\a}, D_{\b]} \bar{\psi}^{(\sigma)}_0(x)\big) \, \psi^{(\sigma)}_0(x) \right], \no
         &=& -\frac{\hbar}{4 m} \, R_{\a\b\c\d}(x)\,u^\b(x) \, \sigma^{\c\d}(x),
\eea
where we have made use of \Eq{44} for the commutator of total covariant derivative
$D_{[\a}, D_{\b]} = \half [D_{\a}, D_{\b}]$. Equation~(\ref{70}) is the generalized \tit{force equation}
for the $v^\a(x)$ congruences and describes the deviation from the classical geodesic motion (described
by $u^\a(x)$) due to the coupling of the particle's spin to the curvature.
The acceleration $a^\a(x) = g^{\a\b}(x) a_\b(x)$ is  force per unit mass
that the spin-$\half$ particle experiences due to the coupling of its spin to its motion.
Due to this non-zero, albeit small, acceleration the
spin-$\half$ particle is no longer in a FFF (\Eq{9}) in which the tetrad is parallel transported along
the particle's geodesic worldline (if one ignores the particle's spin), and is instead more appropriately
described  by a tetrad which is FW transported along the $v^\a(x)$ congruence by \Eq{10}.
In the following sections we explore specific timelike worldlines in the
Schwarzschild metric, which are geodesics if spin is ignored, and the quantum corrections
to the velocity and acceleration given by \Eq{68} and \Eq{70}. We explore the implications of the coupling
of the particle's spin to the gravitational curvature on the Wigner rotation of the particle's spin
and later its effect upon entanglement.

\section{Quantum corrections to circular geodesic motion in the Schwarzschild metric}
We now consider the quantum corrections to some specific classical geodesic motion (when spin is ignored)
in the static spherically symmetric Schwarzschild metric
\be{71}
ds^2 = (1-2M/r)\,dt^2 - \frac{1}{1-2M/r}\,dr^2 - r^2\,( d\theta^2 + \sin^2\theta \,d\phi^2)
\ee
where $x^\a = (t,r,\theta,\phi)$, and
$r_s=2M \equiv 2GM/c^2$ is the Schwarzschild radius of the event horizon produced by the centrally located
gravitating object of mass $M$ (e.g. for the Earth $r_{s\oplus} = 0.89$ cm, and
for the Sun $r_{s\odot} = 2.96$ km).
Since the Schwarzschild metric is independent of the coordinates $t$ (static) and $\phi$
(axial-symmetric) there exist two corresponding Killing vectors $\mbf{\xi} = (1,0,0,0)$ and
$\mbf{\eta} = (0,0,0,1)$, respectively. These Killing vectors can be used to define two integrals of
the geodesic motion, by taking their scalar product with the 4-velocity $\mbf{u}$ tangent to the
geodesic trajectory
\bea{72}
e &\equiv& \mbf{\xi} \cdot \mbf{u} = (1 - 2M/r)\,u^t = (1-2M/r)\frac{dt}{d\tau}, \no
l &\equiv& -\mbf{\eta} \cdot \mbf{u} = r^2 \, \sin^2\theta \, u^\phi = r^2 \, \sin^2\theta \, \frac{d\phi}{d\tau}.
\eea
where $\mbf{u} = (u^t,u^r,u^\theta,u^\phi)$ in a coordinate basis.
The quantity $e$ in \Eq{72} is the total energy per unit rest mass, while $l$ is the orbital angular momentum
per unit rest mass. Both of these quantities remain constant along the geodesic trajectory. As in Newtonian
central force motion, the constancy of the orbital angular momentum allows us to consider, without loss
of generality, geodesic motion in the equatorial plane $\theta = \pi/2$ implying $u^\theta = d\theta/d\tau = 0$.
Writing $u^t$ and $u^\phi$ in terms of $e$ and $l$ respectively, and substituting into the equation for
the normalization of the 4-velocity $\mbf{u}\cdot\mbf{u} = 1$ produces a generalization of the Newtonian
radial energy equation of the form (see Hartle in \cite{tetrads})
\be{73}
\mathcal{E} \equiv \frac{e^2-1}{2} = \half \, \left( \frac{dr}{d\tau} \right)^2 + V_{eff}(r),
\ee
where the effective potential $V_{eff}(r)$ is given by
\be{74}
V_{eff}(r) = \half\,\left[\left(1-\frac{2M}{r}  \right) \, \left(1 + \frac{l^2}{r^2} \right) -1\right]
= -\frac{M}{r} + \frac{l^2}{2 r^2} - \frac{M\, l^2}{r^3}.
\ee
The first two terms of $V_{eff}(r)$ in \Eq{74} are just the central gravitational potential and the
centrifugal barrier potential of Newtonian mechanics. The third term in \Eq{74} represents the
general relativistic correction to Newtonian trajectories, which becomes more pronounced as
$r\to 2M$. Extrema of $V_{eff}(r)$ give rise to stable (+ sign) and unstable (- sign) circular orbits
\be{75}
r_{circ} = \frac{M}{2}\,\left( \frac{l}{M}\right)^2 \, \left(1 \pm \frac{12}{(l/M)^2} \right)^{1/2}.
\ee
The innermost stable circular orbit (ISCO) occurs when $(l/M)^2 = 12$ or $r_{ISCO} = 6M = 3 r_s$.

\subsection{Circular geodesics and geodetic precession of gyroscopes}
For the case of circular geodesics $r=R$, using $u^r = dr/d\tau = 0$ in \Eq{73} along with \Eq{75}
one can deduce the relations
\be{76}
e = \frac{1-2M/R}{\sqrt{1-3M/R}}, \quad l = \frac{\sqrt{M R}}{\sqrt{1-3M/R}}, \quad \trm{on}\spp r=R.
\ee
We define the quantity $\Omega = d\phi/dt$ as the orbital angular momentum of the orbit with respect to
an observer at spatial infinity, where the proper time $\tau$ (the observer's locally measured time) is equal to
the Schwarzschild coordinate time $t$. We then have
\be{77}
\Omega = \frac{d\phi}{dt} = \frac{d\phi/d\tau}{dt/d\tau} = \frac{l/R^2}{e/(1-2M/R)} = \sqrt{\frac{M}{R^3}}.
\ee
Equation~(\ref{77}) is simply the statement of Kepler's third law that the square of the orbital period $P=2\pi/\Omega$
is proportional to the cube of the orbit's radius, $P^2 \propto r^3$.
For a circular geodesic, the observer's 4-velocity takes the form
$\mbf{u} = (0,0,u^\phi,u^t)$, which from \Eq{76} and the normalization condition
$\mbf{u}\cdot\mbf{u} = 1$ yields
\bea{78}
\mbf{u} &=& \left(\frac{\Omega}{\Omega'(r)},0,\,0,\,\frac{\Omega^2}{\Omega'(r)}\right), \\
&=& \left. \left(\frac{1}{\sqrt{1-3M/R}},0,0,\frac{\sqrt{M/R^2}}{\sqrt{1-3M/R}}\right)\right|_{r=R}, \nonumber
\eea
where we have defined
\be{79}
\Omega'(r) \equiv \Omega \, \sqrt{1 - \frac{2M}{r} - r^2\, \Omega^2} \spp \stackrel{r=R}{\longrightarrow}\spp
\Omega \, \sqrt{1 - \frac{3M}{R}}.
\ee
Note that for $M=0$, the frequency $\Omega'(r)$ contains the time dilation
effect $\sqrt{1-V_\phi^2} = \sqrt{1-(r\Omega)^2}$ for a particle executing circular motion with tangential
velocity $V_\phi = r\Omega$ (as measured from spatial infinity).

For each circular orbit defined by $r=R$ simple algebra reveals that the covariant components of the
4-velocity are given by $u_\a = (e,0,0,l)$ which by \Eq{55} shows that the action is given by
$S(x) = -( e\,t + l\,\phi)$ appropriate for a circular orbit derived from the classical Hamilton-Jacobi equation
\Eq{52}.

The interpretation of $\Omega'(r)$ can be inferred from solving for the spatial axes $\bfe{i}(x)$ by treating
the FFF condition \Eq{9} as a set of ordinary differential equations, with $\bfe{0}(x)$ given by \Eq{78}. We
consider the initial conditions such that at $t=0$, $\bfe{r}(x)$ points in the radial direction appropriate for
a stationary observer at $r=R,\phi=0$ (for which $\mbf{u}^{stat}=((1-2M/R)^{-1/2},0,0,0)$), i.e.
$\bfe{r}^{stat}(t=0) = (0,(1-2M/R)^{1/2},0,0)$. The solutions to these equations yields the observer's tetrad
\bea{80}
\lefteqn{\mathbf{e}(x) =
\left[
\begin{array}{c}
  \bfe{t}(x) \\
  \bfe{r}(x) \\
  \bfe{\theta}(x) \\
  \bfe{\phi}(x)
\end{array}
\right] } \no
&=& \left[
\begin{array}{cccc}
   \dfrac{\Omega}{\Omega'(r)} & 0 & 0 & \dfrac{\Omega^2}{\Omega'(r)} \\
   -\dfrac{r\Omega^2}{\sqrt{1-2M/r}\,\Omega'(r)}\,\sin(\Omega'(r)\,t) & \sqrt{1-2M/r} \,\cos(\Omega'(r)\,t) & 0 & -\dfrac{\sqrt{1-2M/r}\,\Omega}{r\Omega'(r)}\,\sin(\Omega'(r)\,t)  \\
  0 & 0 & \dfrac{1}{r} & 0 \\
 \dfrac{r\Omega^2}{\sqrt{1-2M/r}\,\Omega'(r)}\,\cos(\Omega'(r)\,t)  & \sqrt{1-2M/r} \,\sin(\Omega'(r)\,t) & 0 & \dfrac{\sqrt{1-2M/r}\,\Omega}{r\Omega'(r)}\,\cos(\Omega'(r)\,t)
\end{array}
\right] \qquad
\eea

From the first row of \Eq{80} we observe that after one orbital period $t=P=2\pi/\Omega$, the vector $\bfe{r}(P)$
does not return to its initial radial direction $\bfe{r}(0)$. Rather, it is rotated in the direction of
the orbital rotation by an angle
\be{81}
\Delta \phi_{geod} = 2\pi \, \left[\, 1 - \dfrac{\Omega'(r)}{\Omega} \, \right] =
2\pi \, \left[\,1 - \sqrt{1 - 3M/R} \,\right]
\ee
In fact, with respect to the initial, static orthonormal basis $\mathbf{e}(t=0)$ (which is only a FFF frame
when $\Omega'(r)\,t = 2\pi n$)
in which the spatial portions of the
spatial basis vectors $\bfe{r}(0)$, $\bfe{\theta}(0)$ and $\bfe{\phi}(0)$ point directly along
the $r$, $\theta$ and $\phi$ directions, respectively
we have
\Bea
\bfe{r}(t)    &=& \cos(\Omega'(r)\,t) \,\bfe{r}(0) - \sin(\Omega'(r)\,t) \,\bfe{\phi}(0), \\
\bfe{\phi}(t) &=& \sin(\Omega'(r)\,t) \,\bfe{r}(0) + \cos(\Omega'(r)\,t) \,\bfe{\phi}(0).
\Eea
This effect is called (deSitter) geodetic precession. As discussed earlier, each of the spatial axes in \Eq{80}
can be considered as a gyroscope defining a local axis in the observer's laboratory. Equation~(\ref{81}) implies
that these axes precess by an amount $\Delta \phi_{geod}$ per orbit.

It is important to note that we have written the tetrad components in \Eq{80} in terms of $\Omega'(r)$
as a function of $r$ given in \Eq{79}, with $\Omega = \sqrt{M/R^3}$ treated as a constant.
For a specific circular geodesic orbit, $r$ and $R$ are interchangable. However, when computing the
quantum corrections to the velocity \Eq{67} and to the acceleration \Eq{69} we will need to compute
the derivative of the tetrad, especially the variation in $r$. The expression for $\Omega'(r)$ given
in \Eq{79} has the important property that its $r$ derivative evaluated on a given circular geodesic is zero, i.e.
$\pd{r} \Omega'(r)|_{r=R}=0$, upon using the definition of $\Omega$ given in \Eq{77}.


The last point above can be more clearly understood by noting that the
most general form for a  \tit{stationary} metric (i.e $g_{\alpha\beta}(x)$ indpendent
of time, and $g_{0 i}(x)\ne 0$ in general,  indicating a rotation of the spacetime) is
(where we have temporarily reintroduced factors of $c$)
\be{82}
ds^2 = e^{2\Phi/c^2}\left( c\,dt - \frac{1}{c^2} w_i\,dx^i \right)^2 - k_{i j}\,dx^i\,dx^j, \quad i = (1,2,3),
\ee
in the rotating frame or \tit{lattice} coordinates $x^\alpha = (x^0=c t,x^1,x^2,x^3)$.
Here $\Phi$ is \tit{analogous} to the gravitational potential, but now includes length contraction due to the rotation.
The spatial vector $\vec{w} = (w_1,w_2,w_3)$ characterizes the rotation, and the  $3\times 3$ matrix $k_{ij}$ is the
\tit{spatial metric} of the rotating spacetime.

The gravitational field (acceleration) $\vec{a}$ experienced by an
observer with fixed coordinates $\vec{x}=$ constant in the rotating frame has magnitude
\be{83}
|\vec{a}\,| = \left[ k^{ij} \, (\partial_i\,\Phi) \, (\partial _j \Phi) \right]^{1/2},
\ee
where $k^{ij}$ is the inverse matrix of the spatial metric $k_{ij}$.
Further, if we put a gyroscope at $\vec{x}=$ constant in this rotating frame, it will precess at a \tit{proper}
angular velocity $\vec{\Omega}_{gyro}$ with magnitude
\be{84}
|\vec{\Omega}_{gyro}| = \frac{1}{2\sqrt{2}\,c}\, e^{\Phi/c^2}\,
\left[ k^{im}\,k^{jn}\,(\partial_i w_j - \partial_j w_i)\,(\partial_m w_n - \partial_n w_m) \right]^{1/2},
\ee
relative to a freely falling frame (local inertial frame) \cite{rindler}.
The term in the square brackets is just the curved space generalization of
the magnitude of the 3-vector $\trm{curl}\,\vec{w}$ describing the rotation.
In fact, for non-relativistic velocities we have the approximate expressions
\be{85}
\vec{a} = \vec{\nabla} \Phi, \qquad \vec{\Omega}_{gyro} = \frac{1}{2 c} \, \trm{curl}\,\vec{w},
\ee
and the total acceleration experienced by an observer at fixed lattice coordinates $\vec{x}$ is
\be{86}
\vec{a}_{total} = -\vec{a} + 2\,\vec{v}\times\vec{\Omega}_{gyro},
\ee
where $v^i = dx^i/dt$. Thus, the total acceleration experience by an observer at fixed lattice coordinates $\vec{x}$
is the sum of inertial and Coriolis acceleration. In general, $-\vec{a}$ is the acceleration and $-\Omega_{gyro}$ is the
rotation rate of the FFF with respect to the lattice (frame) with coordinates $x^\alpha$.

For the case of the Schwarzschild metric \Eq{71}, we can define (restoring $c=1$)
\be{87}
\phi = \varphi + \Omega\,t, \qquad d\phi = d\varphi + \Omega\,dt,
\ee
where $\Omega$ is the constant angular velocity of a rotating lattice with coordinates $x^{'\alpha}=(t,r,\theta,\varphi')$
with respect to the freely falling frame. After some straightforward algebra, the line element \Eq{71} takes
the form (again for $\theta = \pi/2$)
\be{88}
ds^2 = \left(1 - \dfrac{2 M}{r} - r^2 \Omega^2 \right)
\,\left(dt - \dfrac{r^2 \Omega}{1 -2M/r - r^2 \Omega^2}\,d\varphi' \right)^2
-\dfrac{dr^2}{1-2M/r} - \dfrac{r^2 (1-2M/r)}{1 - 2 M/r - r^2 \Omega^2}\,d\varphi^{'2}.
\ee
From \Eq{83} the points of zero acceleration $|\vec{a}\,|$, corresponding to $\partial_r \Phi(r)=0$ correspond
to a geodesic, i.e. a free particle can remain at rest there (in the new coordinates) and the worldline of the lattice point is a circular geodesic.
A simple calculation yields
\be{89}
\partial_r \Phi(r)|_{r=R}=0 \quad \Rightarrow \quad \Omega^2 = \dfrac{M}{R^3}.
\ee
A calculation of $\Omega_{gyro} \equiv |\vec{\Omega}_{gyro}|$ from \Eq{84} yields
\be{90}
\Omega_{gyro} = \Omega.
\ee
The above is only an apparent coincidence since $\Omega_{gyro}$ is a \tit{proper} rotation rate
relative to the FFF, while $\Omega$ is a coordinate rotation rate (with respect to $t$).
At fixed lattice coordinate $\vec{x}'$ on the circular geodesic $r=R$ we have
$\Delta\tau = (1-3M/R)^{1/2}\,\Delta t \equiv dt/\gamma$ where $\gamma^{-1}(r) = (1 - 2M/r - r^2\Omega^2)$.
Therefore, after one orbital period $\Delta t = 2\pi/\Omega$,
a gyroscope at fixed lattice point $\vec{x}'$ traverses an angle
\be{91}
\varphi_{geo} = \Omega\,\Delta\tau = \Omega\,\frac{\Delta t}{\gamma} =  \Omega \sqrt{1-3M/R}\,\frac{2\pi}{\Omega} =
2\pi\sqrt{1-3M/R}.
\ee
The precession of the gyroscope per orbital revolution relative to original lattice is then
$\Delta\varphi_{geo} = 2\pi - \varphi_{geo}$
which is the same as \Eq{81}.

\subsection{Velocity and acceleration corrections}
To compute the spin-orbit quantum correction to the covariant components of the velocity
\be{92}
v_\a(x) \equiv u_\a(x) + \hbar\,\delta v_\a(x), \qquad
\delta v_\a(x) =  \frac{1}{i\,m} \; \bar{\psi}_0(x) \, \Gamma_\a(x) \, \psi_0(x) + \O(\hbar^2),
\ee
we use \Eq{40} for the spinor connection $\Gamma_\a(x)$, which in turn uses the spin connection $\spincdn{\u}{\ha}{\hb}(x)$
computed from the definition \Eq{40} utilizing \Eq{34} and \Eq{35}.
For $\psi^{(\sigma)}_0(x)$ we form the most general constant component spinor from the positive
energy solutions for spin up and spin down \Eq{60}, in the particle's FFF
\be{93}
\psi_0(x) =
\left(
\begin{array}{c}
  \cos\zeta/2 \\
  e^{i\varphi} \, \sin\zeta/2 \\
  0 \\
  0 \\
\end{array}
\right),
\ee
where $(\zeta,\varphi)$ are the constant polar and azimuthal angles relative to the quantization axis.
In the observer's rest frame the spin points along the direction
$\vec{n}=(\sin\zeta\cos\varphi,\sin\zeta\sin\varphi,\cos\zeta)$.
For motion in the equatorial plane $\theta=\pi/2$, we choose the quantization axis,
the local $\hat{\mathbf{z}}$-axis, to be perpendicular to the orbital plane along $-\bfe{\theta}$, and
hence the local $\hat{\mathbf{x}}$ and $\hat{\mathbf{y}}$ axes to be along $\bfe{r}$ and $\bfe{\phi}$ respectively.

With the spinor \Eq{93}, the quantum velocity correction $\delta v_\a(x)$ has the general form
\be{94}
\delta v_\a(x) = \frac{1}{i\,m}
\left[
\cos^2(\zeta/2)\,\Gamma_\a^{\uparrow\uparrow}(x) + \sin^2(\zeta/2)\,\Gamma_\a^{\downarrow\downarrow}(x)
+ \half \sin\zeta \left(e^{i\varphi}\,\Gamma_\a^{\uparrow\downarrow}(x) + e^{-i\varphi}\,\Gamma_\a^{\downarrow\uparrow}(x)\right)
\right].
\ee
where $\Gamma_\a^{\sigma'\sigma}(x) \equiv \bar{\psi}^{(\sigma')}_0(x) \, \Gamma_\a(x) \, \psi^{(\sigma)}_0(x)$, with
$\psi^{(\sigma)}_0(x)$  for $\sigma',\sigma\in\{\uparrow,\downarrow\}$ the spin up and spin down spinors from \Eq{60}.
For the specific case of the Schwarzschild metric \Eq{71}, the spinor connection has the properties
\Bea
\Gamma_r^{\sigma'\sigma}(x) &=& 0, \hspace{15em} \trm{Scwharzschild metric}\\
\Gamma_\a^{\uparrow\downarrow}(x) &=& \Gamma_\a^{\downarrow\uparrow}(x) = 0, \quad
\Gamma_\a^{\downarrow\downarrow}(x) = - \Gamma_\a^{\uparrow\uparrow}(x), \qquad \a\in\{t,\phi\}, \\
\Gamma_\theta^{\downarrow\downarrow}(x) &=& \Gamma_\theta^{\uparrow\uparrow}(x) = 0, \quad
\Gamma_\theta^{\downarrow\uparrow}(x) = -\left(\Gamma_\theta^{\uparrow\downarrow}(x)\right)^*.
\Eea
A lengthy, but straightforward calculation yields
the correction $\delta v_\a(x)$ for arbitrary radius $r=R$
\be{95}
\delta v_\a(x) =
\left(
\begin{array}{c}
  \delta v_t(x)\\
  \\
  \delta v_r(x) \\
  \\
  \delta v_\theta(x) \\
  \\
  \delta v_\phi(x)
\end{array}
\right)
%
=\left(
\begin{array}{c}
  \displ -\frac{\Omega R \cos\zeta}{2 m \sqrt{1-3 M/R}} \\
  \\
  \displ 0 \\
  \\
  \displ -\frac{\sqrt{1-2M/R}\,\sin(\Omega' t - \varphi)\,\sin\zeta}{2 m R^2} \\
  \\
  \displ \frac{(1-2 M/R) \, \cos\zeta}{2 m R^2 \sqrt{1-3 M/R}} \\
\end{array}
\right),
\ee
where $\Omega = \sqrt{M/R^3}$ and $\Omega' = \Omega\,(1-3 M /R)^{1/2}$.

In general, the velocity correction in \Eq{95} depends on the orientation of the spin
in the FFF. For the case of spin up $(\zeta=0,\varphi=0)$ and spin down $(\zeta=\pi,\varphi=0)$,
$\delta v_\a(x)$ has only $\delta v_\phi(x)$ and $\delta v_t(x)$ components, and thus the motion is along the circular orbit.
For values of $0 < \zeta < \pi$, there is a non-zero $\delta v_\theta(x)$ component, corresponding to
small oscillation in the local $\hat{\mathbf{z}}$ direction at frequency $\Omega'(R)$.


A straightforward calculation of the acceleration $a_\a(x)$ proceeds from \Eq{70}. Here
the Riemann curvature tensor $R_{\a\b\c\d}(x)$ is computed from the Christoffel connection
directly from the metric from \Eq{46} and \Eq{7}, and the world Dirac matrices $\sigma^{\c\d}(x)$
are calculated from the usual Dirac matrices in flat Minkowski spacetime $\sigma^{\hc\hd}$ from
\Eq{65} and \Eq{42}.  This yields
\be{96}
a_\a(x) =
\left(
\begin{array}{c}
  a_t(x) \\
  \\
  a_r(x) \\
  \\
  a_\theta(x) \\
  \\
  a_\phi(x)
\end{array}
\right)
=\hbar\,\left(
\begin{array}{c}
  \displ 0\\
  \\
  \displ -\frac{3\,\Omega^3 \,R\,\cos\zeta}{2 \,m \,(1-3 M/R)} \\
  \\
  \displ -\frac{\Omega \sqrt{1-2 M/R}\,\cos(\Omega'\,t -\varphi)\,\sin\zeta}{2 \,m} \\
  \\
  \displ 0 \\
\end{array}
\right).
\ee
For pure spin up or spin down in the FFF, the acceleration is strictly in the radial direction, corresponding
to the purely circular velocity corrections discussed above.
The acceleration $a_\a(x)$ is the force per unit mass that the spin-$\half$ particle experiences
due to the coupling of its spin to the gravitational curvature. This coupling produces a change of the particle motion
from the geodesic to $\O(1)$, to non-geodesic to $\O(\hbar)$.
With respect to the FFF the acceleration has the components ($\mathbf{a} = a^{\ha}(x)\,\bfe{a}(x)$)
on $r=R$
\be{97}
a^{\ha}(x)=
\left(
\begin{array}{c}
  a^{\hat{t}}(x) \\
  \\
  a^{\hat{r}}(x) \\
  \\
  a^{\hat{\theta}}(x) \\
  \\
  a^{\hat{\phi}}(x)
\end{array}
\right)
= \hbar\,
\left(
\begin{array}{c}
  \displ 0 \\ \\
  \displ \frac{3\,\Omega^3 \,R\,\sqrt{1-2 M/R}\,\cos(\Omega'\,t)\cos\zeta}{2 \,m \,(1-3 M/R)} \\
  \\
  \displ \frac{\Omega \sqrt{1-2 M/R}\,\cos(\Omega'\,t -\varphi)\,\sin\zeta}{2 \,m\,R} \\
  \\
  \displ \frac{3\,\Omega^3 \,R\,\sqrt{1-2 M/R}\,\sin(\Omega'\,t)\cos\zeta}{2 \,m \,(1-3 M/R)}
\end{array}
\right).
\ee

As a useful consistency check, it is worth noting that $a_\a(x)$ arises
directly from the differentiation of $\hbar \delta v_\a(x)$ in \Eq{92} keeping
in mind the action of the covariant derivative on the different type tensorial/spinor quantities.
Thus from \Eq{92} we have
\bea{98}
\lefteqn{\hbar D_\b\delta v_\a(x) =  \frac{\hbar}{m i} \;
D_\b \left( \bar{\psi}^{(\sigma)}_0(x) \, \Gamma_\a(x) \, \psi^{(\sigma)}_0(x)\right) }\no
&=& \frac{\hbar}{m i}
\left(
\, \big(D_\b \bar{\psi}^{(\sigma)}_0(x)\big) \,\Gamma_\a(x) \, \psi^{(\sigma)}_0(x)
+ \bar{\psi}^{(\sigma)}_0(x) \, \nabla_\b\Gamma_\a(x) \, \psi^{(\sigma)}_0(x)
+ \bar{\psi}^{(\sigma)}_0(x) \, \Gamma_\a(x) \, D_\b \psi^{(\sigma)}_0(x)
\right), \no
&=& \frac{\hbar}{m i}
\left(
\, \big(-\bar{\psi}^{(\sigma)}_0(x)\,\Gamma_\b(x)\big) \,\Gamma_\a(x) \, \psi^{(\sigma)}_0(x)
+ \bar{\psi}^{(\sigma)}_0(x) \, \nabla_\b\Gamma_\a(x) \, \psi^{(\sigma)}_0(x)
+ \bar{\psi}^{(\sigma)}_0(x) \, \Gamma_\a(x) \, \Gamma_\b(x) \psi^{(\sigma)}_0(x)
\right), \no
&=&\frac{\hbar}{m i} \bar{\psi}^{(\sigma)}_0(x)
\left(
\nabla_\b \Gamma_\a(x) + [\Gamma_\a(x), \Gamma_\b(x) ]
\right)
\psi^{(\sigma)}_0(x),
\eea
so that upon anti-symmetrization on  the indices $\a$ and $\b$ we have
\be{99}
a_\a(x) = u^\b(x) 2 D_{[\b} \, \delta v_{\a]} =
\frac{\hbar}{m i}\,u^\b(x)\, \bar{\psi}^{(\sigma)}_0(x)
\left(
\nabla_{[\b} \Gamma_{\a]}(x) + [\Gamma_\a(x), \Gamma_\b(x) ]
\right)
\psi^{(\sigma)}_0(x).
\ee
In the above $D_\b$ acting on the spinor connection $\Gamma_\a(x)$ is just the Riemann covariant derivative
as in \Eq{31}, while its action on the spinor $\psi^{(\sigma)}_0(x)$ is given by \Eq{39}.
The term in the parenthesis is just the explicit expression for
$$
[\, D_{\b}, D_{\a} \,] \psi^{(\sigma)}_0(x) =  \spp \frac{i}{4} \, R_{\b\a\c\d}(\Gamma(x))\,\sigma^{\c\d}(x)\, \psi^{(\sigma)}_0(x),
$$
given in \Eq{C6} and hence yields the expression for $a_\a(x)$ in \Eq{70}
(note $R_{\b\a\c\d} = -R_{\a\b\c\d}$).


\subsection{Corrections to the tetrad}
To find the $\O(\hbar)$ corrected tetrad $\bfe{a}^{(\hbar)}(x)$ that defines the FWF (instantaneous, non-rotating
rest frame of the particle) in which the observer detects a null Wigner rotation, one solves the FW transport
equations \Eq{10}, setting $\bfe{0}(x) = \mbf{v}(x)$ given by \Eq{92} and \Eq{93}. Let $\mbf{s}$ be any one of
the three spatial tetrad vectors $\bfe{i}(x)$ with the property $\mbf{s}\cdot\mbf{v}=0$.
We employ a Lindested-Poincar\'{e} perturbation approach \cite{nayfeh}
which allows for a nonlinear frequency correction. The 4-vector $\mbf{s}$ is expanded as
$\mbf{s}(x) = \mbf{s}_0(x) + \hbar\,\mbf{s}_1(x) + \O(\hbar^2)$, and we define a new renormalized time
$\xi \equiv \omega\,\tau = (1 + \hbar\,\omega_1 + \O(\hbar^2))\,\tau$, with $\tau$ the proper time.
The $\O(1)$ equations for $\mbf{s}_0(x)$ reproduce the FFF tetrad of \Eq{80} where the argument of
the sinusoidal functions are $\Omega\,\xi$, reducing to $\Omega\,\tau$ in the limit $\omega_1\to 0$
(which equals $\Omega'(r)\,t$ from the discussion before \Eq{91}).

The condition $\mbf{s}\cdot\mbf{v}=0$ allows one to write $s_0^t(x)$ in terms of $s_0^\phi(x)$ and
$s_1^t(x)$ in terms of both $s_1^\phi(x)$ and $s_0^\phi(x)$ (for components written
with respect to a coordinate basis).
For motion in the equatorial plane
$\theta=\pi/2$ (so that $s_0^\theta(x)=s_1^\theta(x)=0$) the remaining equations for
$s_1^r(x)$ and $s_1^\phi(x)$ can be used to construct a second order equation for $s_1^r(x)$
which takes the form
$$
\frac{d^2 s_1^r(x)}{d\xi^2} + \Omega^2 \,s_1^r(x) = f(\mbf{v},\omega_1)\,s_0^r(x),
$$
where $f(\mbf{v},\omega_1)$ is independent of the time $\xi$, and evaluated on the circular orbit $r=R$.
The trivial particular solution $\mbf{s}_1(x)=0$ can be found by using $\omega_1$ to make $f(\mbf{v},\omega_1)=0$.
A detailed calculation reveals that for an arbitrary circular orbit of radius $r$ and with
$\mbf{v}$ computed for a pure spin up spinor \Eq{60},
\be{100}
\omega_1(r) = \frac{1}{4 M}\,\frac{\Omega^3}{\Omega'(r)}\,\left( 1 + \frac{1}{\Omega'(r)} \right),
\ee
which, in addition, has the property that $\pd{r}\,\omega_1(r)|_{r=R}=0$.
Thus, the first order uniform expansion solution (see \cite{nayfeh}, p125-126) for the FWF tetrad is the
FFF tetrad of \Eq{80} with the frequency $\Omega'(r)$ in the sinusoidal functions replaced by
the renormalized frequency
\be{101}
\Phi'(r) \equiv \big(1 + \hbar\,\omega_1(r)\big)\,\Omega'(r).
\ee
The above implies that the geodetic precession rate per orbit \Eq{91} is increased to
\be{102}
\varphi_{geo}^{(\hbar)} = 2\pi\,\big(1 + \hbar\,\omega_1(r)\big)\,\sqrt{1-3M/R}
\ee
on a circular orbit of radius $r=R$.

The calculation of the spatial tetrad axes $\bfe{i}^{(\hbar)}(x)$, proceeds similarly, but yields complicated
expressions. Most importantly, the correction to the geodetic precession of the spatial axes depends
on the particular orientation $\vec{n}(\zeta,\varphi)$ of the spin in the observer's local frame.
As stated at the beginning of this section, the observer would detect a zero Wigner rotation in
the non-zero acceleration FWF $\bfe{a}^{(\hbar)}(x)$.

\subsection{Wigner rotation}
As discussed in the previous section and at the end of Section IV., an observer in the FWF will detect no Wigner rotation
of the particle's spin. It is instructive to compute the the Wigner rotation as observed from the FFF in which the
classical general relativistic motion is a geodesic when the particle's spin is ignored. The non-trivial portion of
the Wigner rotation matrix $\vartheta^{\hi}_{\,\hj}(x)$ \Eq{26} is calculated using
$p^{\ha}(x) = m\,v^{\ha}(x) = m \,\inveT{a}{\u}(x)\,v^\u(x)$ and $a^{\ha}(x) = m \,\inveT{a}{\u}(x)\,a^\u(x)$ where
$v^\u(x) = u^\u(x) + \hbar \,\delta v^\u(x)$ and $a^\u(x)\equiv \hbar \,\delta a^\u(x)$ are the
$\O(\hbar)$ corrected 4-velocity and 4-acceleration from \Eq{92}, \Eq{93} and \Eq{94},
and $\inveT{a}{\u}(x)$ is the transposed inverse of the FFF tetrad $\e{a}{\u}(x)$ \Eq{80}, of the observer.
The expression for $\chi^{\ha}_{\spp \hb}(x)$ in the non-trivial portion  of
the infinitesimal LT $\lambda^{\ha}_{\spp \hb}(x)$ in \Eq{24}
is obtained from \Eq{23} by replacing $\mbf{u}(x)$ by  $\mbf{v}(x)$, i.e.
$\chi^{\ha}_{\spp \hb}(x) = -\inveT{a}{\u}(x)\,\nabla_{\mathbf{v}}\,\e{b}{\u}(x)$.

Keeping terms to $\O(\hbar)$, a straightforward calculation of \Eq{24} yields
\bea{103}
\lambda^{\ha}_{\spp \hb}(x) &=& \left[ a^{\ha}(x) \, v_{\hb}(x)- v^{\ha}(x) \, a_{\hb}(x) \right]
                              + \chi^{\ha}_{\spp \hb}(x), \no
&=&\hbar \, \left[ \big(\delta a^{\ha}(x)\,u_{\hb}(x) - u^{\ha}(x)\, \delta a_{\hb}(x)\big)
- \inveT{a}{\u}(x)\,\delta v^{\b}(x) \, \nabla_{\b}\,\e{b}{\u}(x) \right], \no
&\equiv& \hbar\,\delta\lambda^{\ha}_{\spp \hb}(x),
\eea
where we have made use the FFF condition \Eq{9} for the tetrad.
The relevant portion of the infinitesimal Wigner transformation from \Eq{26} becomes
\bea{104}
 \vartheta^{\hi}_{\sph \hj}(x) &=& \lambda^{\hi}_{\sp \hj}(x)
 +\frac{\lambda^{\hi}_{\sph \ho}(x)\, v_{\hj}(x) - v^{\hi}(x) \,\lambda_{\hj\,\ho}(x) }{v^{\ho}(x)+1}, \no
&=& \hbar\,\left[\delta\lambda^{\hi}_{\sp \hj}(x)
 +\frac{\delta\lambda^{\hi}_{\sph \ho}(x)\, u_{\hj}(x) - u^{\hi}(x) \,\delta\lambda_{\hj\,\ho}(x) }{u^{\ho}(x)+1} \right].
\eea

In examining the terms in \Eq{102}, we note that since the observer is using the FFF in which $\bfe{0}(x)=\mbf{u}(x)$,
the local components are simply $u^{\ha}(x) = \delta^{\ha}_{\spp \ho}$. By the discussion at the end of Section IV all
the terms involving $u^{\hi}(x)$ in $\lambda^{\hi}_{\sph \hj}(x)$ and $\vartheta^{\hi}_{\sph \hj}(x)$ vanish and
we are left with
$$
\lambda^{\hi}_{\sph \hj}(x) = \chi^{\hi}_{\sph \hj}(x) = \vartheta^{\hi}_{\sph \hj}(x),
$$
with $\chi^{\hi}_{\sph \hj}(x)$ given by the second term in \Eq{101} involving $\delta v^\b(x)$ yielding
$$
\chi^{\hi}_{\sph \hj}(x) = \chi^\uparrow(x) \,
\left(
\begin{array}{ccc}
  0 & \cos(\Omega'\,t) \, \sin(\Omega'\,t)\,\sin\zeta & -\cos\zeta \\
  -\cos(\Omega'\,t) \, \sin(\Omega'\,t)\,\sin\zeta & 0 & -\sin^2(\Omega'\,t)\,\sin\zeta \\
  \cos\zeta & \sin^2(\Omega'\,t)\,\sin\zeta  & 0 \\
\end{array}
\right),
$$
\be{105}
\chi^\uparrow(x) \equiv \hbar\,\frac{(1- 2M/R)}{2\, m\,R^2}.
\ee
In \Eq{105}, $\chi^\uparrow(x)$ is the single, non-zero value
$\chi^{\hat{3}}_{\sph \hat{1}}(x) = -\chi^{\hat{1}}_{\sph \hat{3}}(x)$,
(where $\{\hat{1},\hat{2},\hat{3}\} = $
$\{\hat{\mathbf{x}}, -\hat{\mathbf{z}},  \hat{\mathbf{y}} \} \leftrightarrow$
$\{\bfe{r}, \bfe{\theta}, \bfe{\phi}, \}$)
computed for a pure spin up spinor $(\zeta=0)$, \Eq{60}.
$\chi^\uparrow(x)$ represents a spatial rotation about the local $\hat{\mathbf{z}}$-axis
($-\bfe{\theta}$-axis) perpendicular
to the plane of the orbit, of the FFF with respect to the FWF.

For the orientation we have chosen above for the observer's local axes in the equatorial plane
the infinitesimal spinor rotation matrix \Eq{27} becomes
\be{106}
D^{(1/2)}_{\sigma'\sigma}(W(x); \tilde{\psi}_0) = I + \frac{i}{2}\,\left[ \vartheta_{\hat{2}\hat{3}}(x)\,\sigma_{\hat{1}}
+ \vartheta_{\hat{3}\hat{1}}(x)\,\sigma_{\hat{2}} + \vartheta_{\hat{1}\hat{2}}(x)\,\sigma_{\hat{3}} \right]\, d\tau,
\ee
where
\be{107}
\sigma_{\hat{1}} =
\left(
\begin{array}{cc}
  0 & 1 \\
  1 & 0 \\
\end{array}
\right)\equiv \sigma_{\hat{x}}, \quad
\sigma_{\hat{2}} =
-\left(
\begin{array}{cc}
  1 & 0 \\
  0 & -1 \\
\end{array}
\right)\equiv -\sigma_{\hat{z}}, \quad
\sigma_{\hat{3}} =
\left(
\begin{array}{cc}
  0 & -i \\
  i & 0 \\
\end{array}
\right)\equiv \sigma_{\hat{y}}.
\ee
Here we have defined
\be{108}
\tilde{\psi}_0 =
\left(
\begin{array}{c}
  \cos\zeta/2 \\
  e^{i\varphi}\,\sin\zeta/2 \\
\end{array}
\right) \quad \leftrightarrow \quad \ket{p^{\hi}(x),\sigma}
\ee
as the upper two components of $\psi_0$ of \Eq{93} and which we can associate with the positive energy state
$\ket{p^{\hi}(x),\sigma}$. In \Eq{106} we have indicated that $D^{(1/2)}_{\sigma'\sigma}$ depends upon
the Wigner rotation matrix $W(x)$, which in turns
depends on the zeroth-order spinor wave function $\tilde{\psi}_0$ (which is parallel transported along the circular
geodesic with 4-velocity $\mbf{u}(x)$). For example, if we consider a pure spin up or pure spin down state,
$(\zeta=0,\varphi=0)$ or $(\zeta=\pi,\varphi=0)$ respectively in \Eq{108}, only
$\vartheta_{\hat{3}\hat{1}} = \chi^{\uparrow}(R)\,\cos\zeta$ is non-zero. In these particular cases, $\vartheta_{\hi\hj}(x)$ is
independent of time and hence the time ordering in \Eq{29} can be ignored, and the spinor rotation matrix can
be integrated for finite rotations
\be{109}
D^{(1/2)}_{\sigma'\sigma}(W(x); \tilde{\psi}_0^{(\uparrow,\downarrow)})
= e^{\mp\frac{i}{2}\chi^{\uparrow}(R)\,\tau(R,t)\,\sigma_{\hat{z}}}=
\left(
\begin{array}{cc}
  e^{\mp i\,\chi^{\uparrow}(R)\,\Omega'(R)\,t/2} & 0 \\
  0 & e^{\pm i\,\chi^{\uparrow}(R)\,\Omega'(R)\,t/2} \\
\end{array}
\right),
\ee
where the upper and lower signs are associated with spin up and spin down respectively, and
$\tau(R,t) = \Omega'(R)\,t$. However, for a general spin orientation $0 < \zeta < \pi$ in the FFF \Eq{93}
(e.g. $(\zeta=\pi/2,\varphi=0)$ corresponds to the spinor pointing along $\bfe{r}$,
 $(\zeta=\pi/2,\varphi=\pi/2)$ corresponds to the spinor pointing along $\bfe{\phi}$),
$\vartheta_{\hi\hj}(x)$ is time dependent \Eq{105}, and $D^{(1/2)}_{\sigma'\sigma}(W(x); \tilde{\psi}_0)$ is
not simply a rotation about the local $\hat{\mathbf{z}}$-axis, and can only be computed infinitesimally.
The state $\ket{p^{'\hi}(x'),\sigma} = U(\Lambda(x))\,\ket{p^{\hi}(x),\sigma}
=\sum_{\sigma'}\,D^{(1/2)}_{\sigma'\sigma}(W(x'); \tilde{\psi}_0)\,\ket{p^{\hi}(x'),\sigma'}$
represents the  Wigner rotation of the state $\ket{p^{\hi}(x),\sigma}$ as measured from the FFF at the point $x'$
due to the $\O(\hbar)$ correction of its spin coupling to its motion.

\subsection{Entangled States}
Terashima and Ueda \cite{ueda} considered the Wigner rotation of a spin-singlet state (with the local quantization axis
along $\bfe{\phi}$) created at given point on a non-geodesic equatorial circle, as one of the particles of the bipartite state
circulated the orbit clockwise, and the other counter-clockwise. In this case the observers are taken to be
a set of stationary observers situated around the orbit at $r=R$ (the only non-zero tetrad components are
$\e{r}{r}(x)=(1-2M/r)^{1/2}$, $\e{\theta}{\theta}(x)=1/r$, $\e{\phi}{\phi}(x)=1/r$, and $\e{t}{t}(x)=(1-2M/r)^{-1/2}$).
Since the circular orbit was non-geodesic, $\mbf{a}$
represented the external, non-gravitational 4-acceleration required to keep the particles on the orbit, with each particle
experiencing equal and opposite spatial tangential velocities. Given the velocity and acceleration of each particle,
they calculate the appropriate Wigner rotation and corresponding unitary transformation for each particle and then
apply $D_1(W(\Phi))\otimes D_2(W(-\Phi))$ to the bipartite state. Here the subscripts $1$ and $2$ refer to the
first and second particle comprising the bipartite state, which traverse the orbit from the point of origin $\phi=0$
to $\phi=\Phi(R)$ and $\phi=-\Phi(R)$ respectively.

The result of this calculation is that the spin-singlet state is mixed with the spin-triplet state, and hence
spin measurements in the same direction are not always anti-correlated in the local inertial frames at
$\phi = \pm\Phi$ (i.e. along the local axis $\bfe{\phi}$). The general relativistic effects deteriorates the
perfect anti-correlation in the directions that would be the same as each other if the spacetime were flat \cite{ueda}.
This deterioration is a consequence of the fact that in their scenario
$$
\lambda^{\hi}_{\sph \hj}(x) \neq \chi^{\hi}_{\sph \hj}(x) \neq \vartheta^{\hi}_{\sph \hj}(x),
$$
which results from the imposed accelerations and particular choice of
stationary observers (tetrads) situated around the orbit.

\begin{figure}[h]
\includegraphics[width=5.0in,height=4.0in]{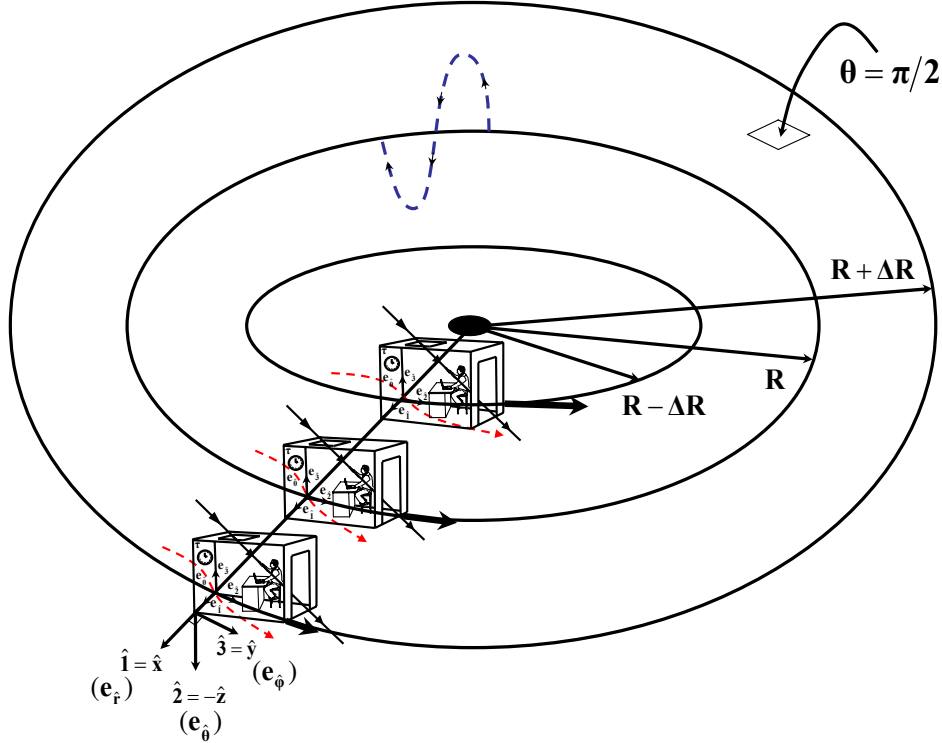}
\caption{(Color online) Bipartite entangled state $\ket{\Psi(x_+,x_-)}$ located on equatorial circular geodesics $r = R\pm\Delta R$
(coordinates $x_\pm$ in text) when the spin of the particle is ignored. We consider the $\O(\hbar)$
spin-orbit coupling of the particle leading to non-geodesic motion on $x_\pm$ (indicated by dotted red lines),
and we consider the Wigner rotation from the geodesic FFF at $r=R$. The dashed sinusoidal blue curve,
perpendicular to the equatorial plane, is the non-geodesic motion when the spins of the particles are not
along the local $\mathbf{z}$-axis ($\bfe{\theta}$). When the spins are oriented parallel or anti-parallel
to the $\mathbf{z}$-axis, the $\O(\hbar)$ velocity corrections are parallel or anti-parallel to the
circular orbit.
}
\label{fig3}
\end{figure}

For the problem that we consider, the coupling of the particle's spin to its motion produces a deviation
of its motion from geodesic to $\O(1)$, to non-geodesic to $\O(\hbar)$. Let us consider the case
of an entangled bipartite state in which to $\O(1)$ the individual particles traverse circular geodesics
at slightly different radii $r=R \pm \delta R$ in the same direction, and then ask how the $\O(\hbar)$
corrections to the particles' orbit effect the bipartite state as observed from a co-circulating FFF situated at $r=R$
(see \Fig{fig3}).
For our entangled state we consider the following normalized bipartite state
\be{110}
\ket{\Psi(x_+,x_-)} = \cos(\Theta/2)\, \ket{\vec{p}(x_+),\uparrow}\,\ket{\vec{p}(x_-),\downarrow}
+ e^{i\Phi}\,\sin(\Theta/2)\, \ket{\vec{p}(x_+),\downarrow}\,\ket{\vec{p}(x_-),\uparrow}.
\ee
In \Eq{110}, $\vec{p}(x)$ represents the spatial portion $p^{\hi}(x)$ of the local momentum $p^{\ha}(x)$ with respect to the
local tetrad $\mathbf{e}(x)$. The argument $x_\pm$ indicates the circular orbits of
radii $r=R \pm \delta R$, respectively. We have also included the arbitrary, constant angles $(\Theta,\Phi)$
in the definition of our bipartite state. For $(\Theta=0,\Phi=\pi/2)$ \Eq{110} is a pure spin-singlet state
($J=0,m_J=0$), and $(\Theta=0,\Phi=0)$ represents the triplet state  ($J=1,m_J=0$),
while for $0 < \Theta < \pi$ it is a linear combination of the spin-singlet and spin-triplet states of zero
magnetic quantum number.

For each particle we define the quantum correction to the velocity as a generalization of \Eq{64} for
a multi-particle state
\bea{111}
v_1^\a(x) &=& \frac{\hbar}{2 m i} \, \left[ \bar{\Psi}(x_+,x_-)\, D_{\a}(x_+)\otimes I \,\Psi(x_+,x_-) - \big(D_{\a}(x_+)\otimes I \,\bar{\Psi}(x_+,x_-)\big) \, \Psi(x_+,x_-) \right], \no
v_2^\a(x) &=& \frac{\hbar}{2 m i} \, \left[ \bar{\Psi}(x_+,x_-)\, I \otimes D_{\a}(x_-) \,\Psi(x_+,x_-) - \big(I \otimes D_{\a}(x_-) \,\bar{\Psi}(x_+,x_-)\big) \, \Psi(x_+,x_-) \right]. \no
& & 
\eea
This leads to expressions analogous to single particle velocity corrections \Eq{68}
\bea{112}
v_{1\a}(x_+) &=& u_\a(x_+) + \frac{\hbar}{m i} \; \bar{\Psi}(x_+,x_-) \, \Big(\Gamma_\a(x_+)\otimes I\Big) \, \Psi(x_+,x_-), \no
v_{2\a}(x_-) &=& u_\a(x_-) + \frac{\hbar}{m i} \; \bar{\Psi}(x_+,x_-) \, \Big(I \otimes \Gamma_\a(x_-)\Big) \, \Psi(x_+,x_-),
\eea
where it is important to note that the entire entangled wave function $\Psi(x_+,x_-)$ \Eq{110} is used to calculate these
local velocity corrections. An explicit calculation yields
\bea{113}
v_{1\a}(x_+) &=& u_\a(x_+) +  \frac{\hbar}{m i}\,\left[\cos^2(\Theta/2)\, \Gamma^{\uparrow\uparrow}_\a(x_+) + \sin^2(\Theta/2)\, \Gamma^{\downarrow\downarrow}_\a(x_+)\right]
= u_\a(x_+) +  \frac{\hbar}{m i}\,\cos\Theta \,\Gamma^{\uparrow\uparrow}_\a(x_+), \no
v_{2\a}(x_-) &=& u_\a(x_-) + \frac{\hbar}{m i}\,\left[\cos^2(\Theta/2) \,\Gamma^{\downarrow\downarrow}_\a(x_-) + \sin^2(\Theta/2) \,\Gamma^{\uparrow\uparrow}_\a(x_-)\right]
= u_\a(x_-) - \frac{\hbar}{m i}\,\cos\Theta\, \Gamma^{\uparrow\uparrow}_\a(x_-), \no
& & 
\eea
where we have used $\Gamma^{\downarrow\downarrow}_\a(x_\pm) = -\Gamma^{\uparrow\uparrow}_\a(x_\pm)$ for the
Schwarzschild metric (section VII.B). Relative to the circular orbit at $r=R$ (which we will indicate
in subsequent expressions with the argument $(x)$ with no subscripts), the $\O(\hbar)$ velocity corrections
\be{114}
\delta v_{\a}(x_\pm) = \pm\frac{\hbar}{m i}\,\cos\Theta \,\Gamma^{\uparrow\uparrow}_\a(x_\pm)
\ee
are parallel or anti-parallel to the direction of motion for the two particles, due to the anti-correlation of the
spins in the bipartite state \Eq{110}. Note that for a pure spin-singlet or spin-triplet entangled
state ($\Theta = \pi/2$, with $\Phi=0$, $\Phi=\pi/2$ respectively) the velocity correction is zero to $\O(\hbar)$,
which implies no $\O(\hbar)$ Wigner rotation correction.

Relative to the FFF for each circular orbit with tangent $u_\a(x_\pm)$, the local 4-velocity components
take the form $u^{\ha}(x_\pm) = \delta^{\ha}_{\spp \ho}$. Thus, the results of section VII.D apply
to each orbit $x_\pm$ and hence
$$
\lambda^{\hi}_{\sph \hj}(x_\pm) = \chi^{\hi}_{\sph \hj}(x_\pm) = \vartheta^{\hi}_{\sph \hj}(x_\pm).
$$
From \Eq{103} $\chi^{\hi}_{\sph \hj}(x_\pm)$ takes the reduced form
\be{115}
\chi^{\hi}_{\sph \hj}(x_\pm) = - \inveT{i}{\u}(x_\pm)\,\delta v^{\b}(x_\pm) \, \nabla_{\b}\,\e{j}{\u}(x_\pm)
\ee
Since $\Gamma^{\uparrow\uparrow}_\a(x_\pm)$ is an expectation computed with a pure spin up state, again
the only non-zero value of the $\chi^{\hi}_{\sph \hj}(x_\pm)$ matrix is
$\chi^{\hat{3}}_{\sph \hat{1}}(x_\pm)|_{\zeta=0} = -\cos\Theta\,\chi^{\uparrow}(x_\pm)$ \Eq{105},
now evaluated at $r=R \pm \delta R$.
Let us define
\be{116}
\vartheta(x_\pm) \equiv \vartheta_{\hat{3}\hat{1}}(x_\pm) = \chi_{\hat{3}\hat{1}}(x_\pm) =
\cos\Theta \,\chi^{\uparrow}(x_\pm).
\ee
As a result of the alternating signs in $\delta v_{\a}(x_\pm)$ \Eq{114}, the Wigner rotation angle
$\vartheta(x_\pm)$ is opposite for the two orbits $x_\pm$.

We now wish to describe the Wigner rotation of the bipartite state as seen from an observer in the FFF at
$r=R$ with 4-velocity $u^\a(x)$. Taking $\delta R/R \ll 1$ we can expand the terms in \Eq{114} and \Eq{115} to
$\O(\delta R)$ as
$$
\inveT{i}{\u}(x_\pm) = \inveT{i}{\u}(x) \pm \delta R \, \frac{\partial}{\partial r}\,\inveT{i}{\u}(x), \quad
\Gamma^{\uparrow\uparrow}_\a(x_\pm) = \Gamma^{\uparrow\uparrow}_\a(x)
 \pm \delta R \, \frac{\partial}{\partial r}\,\Gamma^{\uparrow\uparrow}_\a(x),
$$
to obtain
\bea{117}
\chi^{\hi}_{\sph \hj}(x_\pm) &=& \pm\,\chi^{\hi}_{\sph \hj}(x) + \delta R \, \Delta \chi^{\hi}_{\sph \hj}(x), \no
\chi^{\hi}_{\sph \hj}(x) &=& -\frac{\hbar}{m i}\,\cos\Theta \,\left[
\inveT{i}{\u}(x)\,\Gamma^{\b \uparrow\uparrow}(x) \, \nabla_{\b}\,\e{j}{\u}(x)\right], \\
\Delta \chi^{\hi}_{\sph \hj}(x) &\equiv& -\frac{\hbar}{m i}\,\cos\Theta \,
\left[
 \left(\frac{\partial}{\partial r}\,\inveT{i}{\u}(x)\right) \,\Gamma^{\b \uparrow\uparrow}(x)\, \nabla_{\b}\,\e{j}{\u}(x)
+ \inveT{i}{\u}(x) \,\left(\frac{\partial}{\partial r}\,\Gamma^{\b \uparrow\uparrow}(x)\right)\, \nabla_{\b}\,\e{j}{\u}(x) \right. \no
& & \left. \hspace{6em}
+ \, \inveT{i}{\u}(x) \,\Gamma^{\b \uparrow\uparrow}(x) \frac{\partial}{\partial r}\,\left(\nabla_{\b}\,\e{j}{\u}(x)\right)
\right], \nonumber
\eea
where $\Gamma^{\b \uparrow\uparrow}(x) = g^{\b\a}(x)\,\Gamma^{\uparrow\uparrow}_\a(x)$.
Thus, \Eq{116} becomes
\be{118}
\vartheta(x_\pm) = \pm\vartheta(x) + \Delta\vartheta(x), \quad
\vartheta(x) = \cos\Theta \,\chi^{\uparrow}(x), \quad
\delta\vartheta(x) \equiv \delta R \, \Delta \chi_{\hat{3}\hat{1}}(x)
\ee
In \Eq{117} and \Eq{118} the leading $\pm$ in front of $\chi^{\hi}_{\sph \hj}(x)$ and $\vartheta(x)$ arises
directly from the alternating sign in \Eq{114} for the particles on orbits of radii $r=R \pm \delta R$.
The additional $\pm$ sign, resulting from expansion of the argument $x_\pm$,
combines with former $\pm$ to produce the $\O(\delta R)$ with the same sign for both particles.

From the FFF at radius $r=R$ the unitary transformation of the bipartite state \Eq{110} is
\bea{119}
\lefteqn{D[W\big(\vartheta(x) + \Delta\vartheta(x)\big)] \otimes D[W\big(-\vartheta(x) + \Delta\vartheta(x)\big)] =} \no
& &
\left[
\begin{array}{cc}
  e^{-i/2\big(\vartheta(x) + \Delta\vartheta(x)\big)d\tau} & 0 \\
  0 & e^{i/2\big(\vartheta(x) + \Delta\vartheta(x)\big)d\tau} \\
\end{array}
\right] \,
\otimes
\left[
\begin{array}{cc}
  e^{-i/2\big(-\vartheta(x) + \Delta\vartheta(x)\big)d\tau} & 0 \\
  0 & e^{i/2\big(-\vartheta(x) + \Delta\vartheta(x)\big)d\tau} \\
\end{array}
\right].\quad\qquad
\eea
The action of \Eq{119} upon the bipartite state \Eq{120} produces the Wigner rotated state
\be{120}
\ket{\Psi'(x)} = \cos(\Theta/2)\, e^{-i\,\vartheta(x)d\tau}\,\ket{\vec{p}(x),\uparrow}\,\ket{\vec{p}(x),\downarrow}
+ e^{i\Phi}\,\sin(\Theta/2)\, e^{i\,\vartheta(x)d\tau}\,\ket{\vec{p}(x),\downarrow}\,\ket{\vec{p}(x),\uparrow},
\ee
as observed from a FFF observer circulating the geodesic circular orbit at $r=R$ with 4-velocity $u^\a(x)$.
Note that to lowest order the  $\O(\delta R)$ have cancelled identically. Further,
for the pure spin-singlet $(\Theta=\pi/2,\Phi=0)$ and pure spin-triplet state $(\Theta=\pi/2,\Phi=\pi/2)$
of zero magnetic quantum number, there is no observed Wigner rotation, since
from \Eq{118} $\vartheta(x) = \cos\Theta \,\chi^{\uparrow}(x) \,|_{(\Theta=\pi/2)}= 0$.

For any other observer with tetrad $\mathbf{e}'(x)$, instantaneously coincident at $r=R$ with
the FFF observer with tetrad $\mathbf{e}(x)$, the observed Wigner rotation of the bipartite state
is much more complicated. It can be obtained by finding the components of the 4-velocity $\mbf{v}(x)$ and
4-acceleration $\mbf{a}(x)$ for each particle relative to the new observer's tetrad, which is calculated
by a local (position dependent) LT between the two instantaneously coincident frames,
$\bfe{a}'(x) = \Lambda_{\ha}^{\spp \hb}(x) \, \bfe{b}(x)$.
Finally, by using the first lines of \Eq{103} and \Eq{104} one can work out the  Wigner rotation matrix
as observed from the new frame.

\subsection{Comments on Spin-Momentum Entanglement and Wigner Rotation}
A general two particle state takes the form
\bea{121}
\ket{\Psi(x_1,x_2)} &=& \sum_{\sigma_1,\sigma_2}\int \int \tilde{d} p_1 \tilde{d} p_2 \,
     g_{\sigma_1\sigma_2}(\vec{p}_1,\vec{p}_2)
  \ket{\vec{p}_1(x_1),\sigma_1} \, \ket{\vec{p}_2(x_2),\sigma_2}, \no
& & \sum_{\sigma_1,\sigma_2}\int \int \tilde{d} p_1 \tilde{d} p_2 \, |g_{\sigma_1\sigma_2}(\vec{p}_1,\vec{p}_2)|^2=1, \no
\tilde{d}p &=& \frac{1}{(2\pi)3}\,\theta(p^{\hat{0}})\, \delta^{(3)}(\vec{p}-\vec{p}')\,\delta^{\sigma'}_\sigma
= \frac{d^3p}{(2\pi)^3 2 p^{\hat{0}}},
\eea
where $g_{\sigma_1\sigma_2}(\vec{p}_1,\vec{p}_2)$ is the joint spin-momentum distribution function and
$\tilde{d}p$ is the (local) Lorentz invariant integration measure. From \Eq{14}, an infinitesimal LLT
$\mathcal{U} = U(\Lambda(x_1))\otimes U(\Lambda(x_2))$ of the state $\ket{\Psi(x_1,x_2)}$ will mix
spin $\sigma$ and 4-momentum $\m{p}$ since the Wigner rotation angle is momentum dependent,
the same as the flat spacetime result discussed by Gingrich and Adami \cite{adami_spin}.
As a result, the reduced two particle spin density matrix formed by tracing out the momentum
of the particles will exhibit an observer  (Lorentz transformation) dependent
entanglement, measured e.g. by  Wootter's  concurrence.

The results of the previous sections indicate that, due to the spin-curvature coupling of the massive particle in CST
resulting from the particle's orbit (4-velocity) being determined from its Dirac current,
the Wigner rotation also depends on the initial orientation of the particle's spin in its local frame
(see the wavefunction \Eq{93}, the velocity \Eq{95} and acceleration \Eq{96} corrections, and subsequent expressions).
Thus, in CST the observed Wigner rotation is
a function of not only of the LLT $\Lambda$ and 4-momentum $\m{p}$, but also the spin orientation
$\vec{n}(\zeta,\varphi)$ in the local (laboratory) frame, which we indicate as
\be{122}
W = W(\Lambda,\vec{p},\vec{n}), \qquad \trm{in CST}.
\ee

\section{Radially infalling geodesic motion in the Schwarzschild metric}
To explicitly demonstrate the local spin orientation dependence of the Wigner rotation \Eq{122}, we consider
the simpler case of the quantum corrections to the FFF of a radially infalling particle, and the $\O(\hbar)$
corrected tetrad in which the Wigner rotation would be measured null. For a particle dropped in from spatial infinity
in the equatorial plane ($\theta=\pi/2$)
with zero velocity and zero orbital angular momentum ($e=1,l=0$), the radial energy equation \Eq{73}
with the ansatz $\mbf{u} = (u^t, u^r, 0, 0)$ yields $\mbf{u} =(1/(1-2M/r), -(2M/r)^{1/2},0,0)$,
where the minus sign in $u^r$ indicates a radially \tit{infalling} geodesic.
The freely falling frame tetrad (FFF) in the equatorial plane is given by
\be{123}
\mathbf{e}(x) =
\left[
\begin{array}{c}
  \bfe{t}(x) \\ \\
  \bfe{r}(x) \\ \\
  \bfe{\theta}(x) \\ \\
  \bfe{\phi}(x)
\end{array}
\right]
= \left[
\begin{array}{cccc}
   \dfrac{1}{1-2M/r} & \,-\sqrt{2M/r} & 0 & 0 \\ \\

   -\dfrac{\sqrt{2M/r}}{1-2M/r} & 1 & 0 & 0  \\\\
  0 & 0 & 1/r & 0 \\ \\
   0 & 0 & 0 & 1/r
\end{array}
\right]
\equiv
\left[
\begin{array}{c}
  \bfe{0}(x) \\ \\
  \bfe{1}(x) \\ \\
  \bfe{2}(x) \\ \\
  \bfe{3}(x)
\end{array}
\right],
\ee
which satisfies the FFF condition \Eq{9}, $\nabla_{\mathbf{u}} \, \bfe{a}(x)=0$, where $\bfe{0}(x) \equiv \mbf{u}(x)$,
and the orthonormalization condition $\mathbf{e}(x)\cdot\mathbf{g}(x)\cdot\mathbf{e}^T(x)=\mbf{\eta}$.
(Note, we have used the same local axes designation as in \Fig{fig3}).

\begin{figure}[h]
\includegraphics[width=3.75in,height=2.75in]{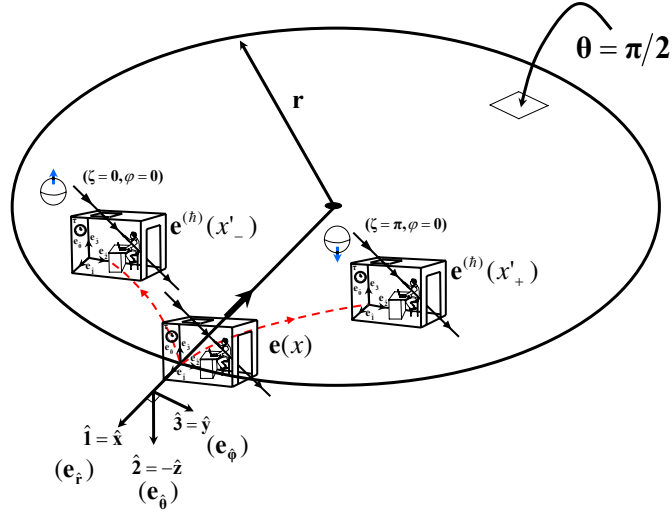}
\caption{(Color online)
Bipartite entangled singlet state $\ket{\Psi(x,x)}_{12}$ \Eq{127} on a radially infalling geodesic with
FFF $\mathbf{e}(x)$ with 4-velocity $\bfe{0}(x) = \mbf{u}(x)$ \Eq{123}. The coupling of the spin to
the spacetime curvature creates a spin dependent deflection of the particles with the
new 4-velocity (Dirac current) $\mbf{v}(x)=\mbf{u}(x)+\hbar\,\delta\mbf{v}(x)$,
with $\delta v_\phi(x) = -\cos\zeta / (4 M r^2)$ \Eq{124}, which is along the local $\mp\mathbf{y}$-axis
for pure spin up $(\zeta=0,\varphi=0)$ and spin down $(\zeta=\pi,\varphi=0)$ orientations
(where $\vec{n} = (\sin\zeta\cos\varphi,\sin\zeta\sin\varphi,\cos\zeta)$ is the spin orientation
in the local inertial frame at $x$).
Under a LLT, the state $U(\Lambda)\ket{\Psi(x,x)}_{12} = \ket{\Psi(x'_+,x'_-)}_{12}$ at the
transformed points $\mbf{x}\to \mbf{x}'_\pm = \mbf{x} + \mbf{v}\,d\tau$
remains a spin singlet state if the two observers use
\tit{different} $\O(\hbar)$-corrected FFF tetrads (since $\delta\mbf{a}(x)=0$),
$\mathbf{e}^{(\hbar)}(x'_-;\zeta=0,\varphi=0)$, for spin up at $x_-$, and
$\mathbf{e}^{(\hbar)}(x'_+;\zeta=\pi,\varphi=0)$ for spin down at $x_+$.
}
\label{fig4}
\end{figure}

Using the general spin orientation in the FFF given by \Eq{93},
corresponding to a local spin direction of $\vec{n} = (\sin\zeta\cos\varphi,\sin\zeta\sin\varphi,\cos\zeta)$,
we obtain velocity correction $\delta v^\a(x)$  \Eq{94} and the acceleration correction $\delta a^\a(x)$ from \Eq{99}
\bea{124}
v^\a(x) &=&  u^\a(x) + \hbar \, \delta v^\a(x) =
\left(
\begin{array}{c}
   v_t(x)\\
  \\
  v_r(x) \\
  \\
  v_\theta(x) \\
  \\
  v_\phi(x)
\end{array}
\right)
=\left(
\begin{array}{c}
  \displ 1/(1-2 M/r) \\
  \\
  \displ -\sqrt{2M/r} \\
  \\
  \displ -\hbar\sin\zeta \,\sin\varphi / (4 M r^2) \\
  \\
  \displ -\hbar\cos\zeta / (4 M r^2) \\
\end{array}
\right),   \no
\delta a^\a(x) &=& 0.
\eea
For a spin orientation in the FFF in the local $\mathbf{x}$-$\mathbf{z}$  ($\hat{r}$-$\hat{\theta}$) plane in
which $\varphi = 0$, the effect of the velocity correction is a deflection along the local $\mathbf{y}$ ($\hat{\phi}$)
direction in the equatorial plane by $\delta v_\phi(x) = -\cos\zeta / (4 M r^2)$.
For $\sin\zeta\sin\varphi\ne 0\,(n_y\ne 0)$ the velocity correction takes the motion of the particle
out of the equatorial plane.
For pure spin up/spin down
($\zeta = (0,\pi)$, respectively) we have $\delta v_\phi^{\uparrow\,\downarrow}(x) = \mp\hbar / (4 M r^2)$,
as depicted in \Fig{fig4}.
Since the acceleration correction is zero in \Eq{124}, the Wigner rotation
is of the form given in \Eq{115} for an arbitrary tetrad $\mathbf{e}(x)$, and arises solely from the rotation
of the tetrad $\chi^{\hi}_{\; \hj}(x)$ and the velocity correction $\delta v^\a(x)$. Since
$\delta a^\a(x)=0$, we can solve for an $\O(\hbar)$ corrected tetrad $\mathbf{e}^{(\hbar)}(x)$ which is
the FFF in which the observer measures a zero Wigner rotation.
Solving
\be{125}
\trm{FFF in which W = 0}: \hspace*{0.5in}
\nabla_{\mathbf{v}} \, \bfe{a}^{(\hbar)}(x)=0, \qquad \bfe{0}^{(\hbar)}(x) \equiv \mbf{v}(x),
\ee
yields
\be{126}
\mathbf{e}^{(\hbar)}(x) =
\left[
\begin{array}{c}
  \bfe{t}^{(\hbar)}(x) \\ \\
  \bfe{r}^{(\hbar)}(x) \\ \\
  \bfe{\theta}^{(\hbar)}(x) \\ \\
  \bfe{\phi}^{(\hbar)}(x)
\end{array}
\right]
= \left[
\begin{array}{cccc}
   \dfrac{1}{1-2M/r} & \,-\sqrt{2M/r} & \,-\hbar\,\dfrac{\sin\zeta\,\sin\varphi}{4 M r^2} & \,-\hbar\,\dfrac{\cos\zeta}{4 M r^2} \\ \\
%
   -\dfrac{\sqrt{2M/r}}{1-2M/r} & 1 & \,\hbar\,\dfrac{\sin\zeta\,\sin\varphi}{\sqrt{8 M^3 r^3}} & \,\hbar\,\dfrac{\cos\zeta}{\sqrt{8 M^3 r^3}}  \\\\
%
  \hbar\,\dfrac{\sin\zeta \sin\varphi}{4 M r (1 - 2 M/r)}  & -\hbar\,(r-M)\,\dfrac{\sin\zeta\,\sin\varphi}{\sqrt{8 M^3 r^3}} & 1/r & 0 \\ \\
%
   \hbar\,\dfrac{\cos\zeta}{4 M r (1 - 2 M/r)} & \hbar\,\dfrac{\cos\zeta}{\sqrt{8 M^3 r^3}} & 0 & 1/r\\
\end{array}
\right],
\ee
which satisfies the normalization condition
$\mathbf{e}^{(\hbar)}(x)\cdot\mathbf{g}(x)\cdot\mathbf{e}^{(\hbar)T}(x)=\mbf{\eta} + \O(\hbar^2)$.
In \Eq{125} we see explicitly that the FFF $\mathbf{e}^{(\hbar)}(x')=\mathbf{e}^{(\hbar)}(x';\zeta,\varphi)$
in which $W=0$ at the transformed point $x'$ depends on the spin orientation $\vec{n}(\zeta,\varphi)$ at $x$.

An implication for entangled states is as follows. Consider, for example, the spin singlet state
at the spacetime point $x$ (see \Eq{110})
\be{127}
\ket{\Psi(x,x)}_{12} =  \ket{\vec{p}(x),\uparrow}_1\,\ket{\vec{p}(x),\downarrow}_2
- \ket{\vec{p}(x),\downarrow}_1\,\ket{\vec{p}(x),\uparrow}_2.
\ee
Since $\delta v_\phi^{\uparrow\,\downarrow}(x) = \mp 1 / (4 M r^2)$, the spin up states will be
deflected along the local $-\mathbf{y}$-axis, while the spin up particles will be deflected along
the local $+\mathbf{y}$-axis.
In order for the state to be observed with zero Wigner rotation, i.e. as the state
\be{128}
\ket{\Psi(x'_+,x'_-)}_{12} =  \ket{\vec{p}(x'_-),\uparrow}_1\,\ket{\vec{p}(x'_+),\downarrow}_2
- \ket{\vec{p}(x'_+),\downarrow}_1\,\ket{\vec{p}(x'_-),\uparrow}_2.
\ee
where $x'_\pm$ indicates the transformed particle positions with small displacements along the local $\pm\mathbf{y}$-axis,
we need the two local Lorentz observers (or local inertial frames, LIF)  to be using \tit{different} tetrads
$\mathbf{e}^{(\hbar)}$ at $x'_\pm$, i.e. the LIF at $x_-$ using the tetrad appropriate
for spin up $\mathbf{e}^{(\hbar)}(x'_-;\zeta=0,\varphi=0)$, and the LIF at $x_+$ using
the tetrad appropriate for spin down $\mathbf{e}^{(\hbar)}(x'_+;\zeta=\pi,\varphi=0)$, see \Fig{fig4}. We cannot
use a \tit{single} class of observers throughout spacetime, since $\mathbf{e}^{(\hbar)}(x')$ depends
on the initial spin orientation $\vec{n}(\zeta,\varphi)$ at $x$.
For a pair of observers other than $\mathbf{e}^{(\hbar)}(x';\zeta=\{\pi,0\},\varphi=0)$ at $x'_\pm$,
the general LLT \Eq{28a} and \Eq{28b} indicates that transformed state $U(\Lambda(x))\,\ket{\Psi(x,x)}_{12}$
would be an entangled state composed primarily of a singlet state, with an $\O(\hbar)$ superposition
of a triplet state. Such observer's would detect an apparent decrease in the maximum EPR correlation
for the two particles at $x'_\pm$, while those using $\mathbf{e}^{(\hbar)}(x';\zeta=\{\pi,0\},\varphi=0)$
would not.

\section{Summary and Conclusions}
In flat spacetime the positive energy, single particle state $\ket{\vec{p},\sigma}$ of a massive particle is given
by its momentum, and spin-$j$ components $\sigma$
 along some quantization axis. Under a LT $\Lambda$, describing the same
flat spacetime from an inertial reference frame moving with constant velocity (zero acceleration)
relative to the original inertial frame, the state is transformed by \Eq{11}, in which $W(\Lambda,\vec{p})$
is the Wigner rotation angle of the $(2j+1) \times (2j+1)$ $O(3)$ rotation matrix,
which mixes up the spin components of the transformed state. If we consider wavepacket states composed of an integration
of single particle states over a momentum distribution, then the momentum dependent Wigner rotation angle
$W(\Lambda,\vec{p})$ will transform each momentum component differently. For a general wavepacket
comprised of a joint distribution over two particle states, the unitary transformation in \Eq{11} will
lead to a spin-momentum entanglement, such that if we consider the reduced spin density matrix
of a bipartite state, there will be degradation of spin-spin entanglement.

In curved spacetime, we replace the positive energy, single particle state of flat spacetime
with the local state $\ket{\vec{p}(x),\sigma}$ valid in the locally flat Lorentz tangent plane to the
the CST at the point $x$. By the equivalence principle, the laws of SR hold in this tangent plane.
Measurements of properties of this state are made from the reference frame of a massive observer
with 4-velocity $\m{u}_{obs}(x)$, instantaneously collocated at the spacetime point $x$.
We define this observer by the orthonormal tetrad $\bfe{a}(x)$
which comprise the four axes of his local laboratory. The three axes $\bfe{i}(x)$
comprise the spatial axes at the origin of the observer's local laboratory, and
the temporal axis $\bfe{0}(x)=\m{u}_{obs}(x)$, tangent to the observer's worldline, determines
the local rate at which his clock ticks. The observer measures the local components $p^{\ha}(x)$
of the 4-momentum $\m{p}(x) = m \m{u}(x)$ of a particle passing through his local laboratory at $x$
by projecting $\m{p}(x)$ onto the tetrad via $p^{\ha}(x) = \inveT{a}{\a}(x)\,p^\a(x)$, where
$E(x)=p^{\ho}(x)$ is energy of the passing particle as measured by the observer, and
$p^{\hi}(x)$ are locally measured 3-momentum components. Under a local Lorentz transformation $\Lambda(x)$,
which transforms between observers in different states of motion collocated at $x$ (i.e. stationary,
freely falling, circular orbit, or under arbitrary acceleration), the state $\ket{\vec{p}(x),\sigma}$
transforms under a local Wigner rotation $W(\Lambda(x),\vec{p}(x))$ \Eq{14}, which generalizes
the flat spacetime result of \Eq{11} by the equivalence principle. The same SR considerations of
spin-momentum entanglement for wavepackets states hold as well in CST, but now locally at each spacetime point $x$.

In this work we have considered massive particles of spin $\half$. Since the observed particle is massive, we can
always find a non-rotating, (in general) accelerating observer that is instantaneously at rest with the particle,
called a Fermi-Walker frame (FWF). If the particle follows a geodesic (zero acceleration) the FWF reduces
to the particle's freely falling frame (FFF).
We have shown that in the FWF, the Wigner rotation angle is null. In any other
frame, the observer would measure a non-zero Wigner rotation for the spin of the particle.

It is a postulate of General Relativity that the force free motion of particles are geodesics, i.e.
trajectories for which the acceleration is zero. This is true if the particle is assumed to possess no spin.
If the particle does possess spin (even classically), the spin of the particle couples to the curvature
and creates non-geodesic motion. In this work, we have considered the motion of a quantum spin $\half$  particle,
as determined by its Dirac current, leading to non-geodesic motion to first order in $\hbar$.
We have found that the momentum (and acceleration) corrections depend on the initial spin orientation
of the particle in its local frame, and hence the momentum dependent Wigner rotation will transform
each of these spin orientations differently. Whereas in flat spacetime the Wigner rotation angle $W$ depends
on both the Lorentz transformation $\Lambda$ and the particle's momentum $\vec{p}$, in CST $W$ depends on the local
analogues of these quantities, as well as on the spin orientation $\vec{n}$ of the particle in
its local frame. In addition, we have explored
the evolution of the entanglement of a bipartite state on two infinitesimally close circular orbits,
as observed from the FFF of circular orbit at the averaged radius. We have also shown that an entangled Bell
state formed at the spacetime point $x$ for a pair of collocated freely falling observers in the equatorial
Schwarzschild plane,  will be observed to be in the same entangled state, i.e. observed to
have zero Wigner rotation under a local Lorentz transformation, if at the spatially separated transformed
spacetime points, it is observed by two different $\O(\hbar)$-corrected freely falling observers.
For any other pair of observers a non-zero Wigner rotation would be measured, which could lead
to an apparent decrease in the measured EPR correlation between the two particles.

\begin{acknowledgments}
PMA wishes to acknowledge the support of  the Air Force Office of Scientific Research (AFOSR) for this work.
\end{acknowledgments}

\appendix

\section{Derivation of the local infinitesimal Wigner rotation \Eq{26}}\label{appA}
In this appendix we derive the infinitesimal form of the Wigner rotation to $\O(d\tau)$
in the observer's LIF given in \Eq{25} and \Eq{26},
\be{A1}
W^{\ha}_{\spp \hb}(x) \equiv \delta^{\ha}_{\spp \hb} + \vartheta^{\ha}_{\spp \hb}(x)\, d\tau,
\ee
where
\bea{A2}
 \vartheta^{\ho}_{\sph \ho}(x) &=& \vartheta^{\ho}_{\sph \hi}(x) = \vartheta^{\hi}_{\sph \ho}(x) = 0, \no
 \vartheta^{\hi}_{\sph \hj}(x) &=& \lambda^{\hi}_{\sp \hj}(x)
 +\frac{\lambda^{\hi}_{\sph \ho}(x)\, p_{\hj}(x) - p^{\hi}(x) \,\lambda_{\hj\,\ho}(x) }{p^{\ho}(x)+m},
\eea
from the formal definition of the Wigner rotation in \Eq{15}
\be{A3}
 W^{\ha}_{\spp \hb}(x) \equiv  \left[ L^{-1}(\Lambda p(x)) \cdot \Lambda(x) \cdot L(p(x)) \right]^{\ha}_{\spp \hb}.
\ee

As the particle moves in curved spacetime from a point from $x^\a \to x^{'\a} = x^\a + u^\a(x) d\tau$.
its 4-momentum  undergoes a local Lorentz transformation (LLT)
$p^{\ha}(x) \to p^{'\ha}(x)= \Lambda^{\ha}_{\spp \hb}(x) \,p^{\hb}(x) = p^{\ha}(x) + \d p^{\ha}(x)$
where
\be{A4}
\Lambda^{\ha}_{\sph \hb}(x) = \delta^{\ha}_{\spp \hb} + \lambda^{\ha}_{\sph \hb}(x)\,d\tau,
\ee
and
\be{A5}
\delta p^{\ha}(x) = \lambda^{\ha}_{\spp \hb}(x) \,p^{\hb}(x)\, d\tau.
\ee
The exact form of the infinitesimal LLT $\lambda^{\ha}_{\sph \hb}(x)$ is given in \Eq{24}. However,
in the subsequent calculations  we will only need that fact that $\lambda_{\ha\hb}(x) = \eta_{\ha\hc}\,\lambda^{\hc}_{\sph \hb}(x)$
is anti-symmetric in its lower two indices
\be{A6}
\lambda_{\ha\hb}(x) = -\lambda_{\hb\ha}(x).
\ee
Note that in the observer's LIF, indices are raised and lowered with the flat spacetime metric
$\eta_{\ha\hc} =$ diag$(1,-1,-1,-1)$. The dot product of two vectors $p^{\ha} = (p^{\ho},\vec{p})$
and $q^{\ha} = (q^{\ho},\vec{q})$ is given by
$p^{\ha}(x)\,q_{\ha}(x) = p^{\ho}(x)\, q^{\ho}(x) -\vec{p}(x)~\cdot~\vec{q}(x) $ where
$\vec{p}(x)\cdot\vec{q}(x) \equiv \d_{\hi\hj} \,p^{\hi}(x) \,q^{\hj}(x)$.
Further, because the 4-momentum is normalized to $p^\a(x) \,p_\a(x) = m^2$, its time component $p^{\ho}(x)$
is determined by it spatial 3-momentum components $ p^{\ho}(x) = \sqrt{\vec{p}^{\;2}(x) + m^2}$.
In the following we will drop the argument $(x)$ on all spacetime dependent quantities for readability.

In the particle's rest frame we define the \tit{standard momentum} $k^{\ha} \equiv (1,0,0,0)$.
The interpretation of the Wigner rotation $W$ in \Eq{A1} is that \tit{standard boost} $L(\vec{p})$ takes the standard
momentum $k^{\ha} \to p^{\ha}$ with 3-momentum components $\vec{p}$, while the arbitrary LLT $\Lambda$ takes
$p^{\ha}\to p^{'\ha} = \Lambda^{\ha}_{\spp \hb}\,p^{\hb}$ with 3-momentum components $\vec{p}\,'$.
Finally, $L^{-1}(\vec{p}^{\sp '})$ takes
$p^{'\ha}\to k^{\ha}$, i.e. back to the rest momentum, which in general can differ from
the standard momentum $k^{\ha}$ by at most a spatial rotation $k^{\ha} = W^{\ha}_{\spp \hb}\,k^{\hb}$.
Thus, Wigner's \tit{little group}, or the invariant subgroup of the massive particle's rest frame is
$O(3)$, i.e. spatial rotations.

The calculation of the infinitesimal form of the Wigner rotation will proceed by expanding \Eq{A3} to
$\O(d\tau)$ using the infinitesimal forms of each of the component matrices. The infinitesimal form
of the arbitrary LLT $\Lambda$ is given by \Eq{A4}, so our first goal is to find the infinitesimal form of the
standard boosts $L(\vec{p})$ and $L^{-1}(\vec{p}^{\sp '})$.  The form of the standard boost is given by \Eq{13}
(with $p^\u \to p^{\ha}(x)$)
which we repeat below
\bea{A7}
 L^{\ho}_{\spp \ho}(\vec{p}) &=& \gamma = \frac{p^{\ho}}{m}\no
 L^{\hi}_{\spp \ho}(\vec{p}) &=& \frac{p^{\hi}}{m}, \quad L^{\ho}_{\spp \hi}(\vec{p}) = -\frac{p_{\hi}}{m}, \no
 L^{\hi}_{\sp \hj}(\vec{p}) &=& \delta^i_{\sp j} - (\gamma-1)\, \frac{p^{\hi} p_{\hj}}{|\vec{p}\,|^2}, \qquad i,j = (1,2,3),
\eea
where $\gamma = p^{\ho}/m = E/m \equiv e$ is the particle's energy per unit rest mass.
Note that for the flat spacetime metric $\eta_{\ha\hb} =$diag$(1,-1,-1,-1)$,
$p_{\hat{0}} = p^{\hat{0}}$ and $p_{\hi} = -p^{\hi}$ which then agrees with the expressions in \cite{weinberg_qft1}.

To find $L^{-1}(\vec{p}^{\sp '})$, note that $L^{-1}(\vec{p}) = L(-\vec{p})$, i.e. the inverse of
the standard boost $L(\vec{p})$ which takes $k^{\ha} \to p^{\ha}$ is simply the standard
boost in the anti-parallel direction along $-\vec{p}$. However, under the inversion $\vec{p}\to -\vec{p}$,
only the $L^{\ho}_{\spp \hi}$ and $L^{\hi}_{\spp \ho}$ terms in \Eq{A7} which are linear in $\vec{p}$ change sign, while
$L^{\ho}_{\spp \ho}$ and $L^{\hi}_{\sp \hj} $ do not. As such, let us first consider $L(\vec{p}^{\sp '})$, and
then develop $L^{-1}(\vec{p}^{\sp '})$ by changing the sign of the argument of the former.

With the substitution $p^{\ha} \to p^{' \ha} = p^{\ha} + \d p^{\ha}$ in the first two lines of \Eq{A7} we have
\bea{A8}
 L^{\ho}_{\spp \ho}(\vec{p}^{\sp '}) &=&  \frac{p^{\ho}}{m} + \frac{\d p^{\ho}}{m}\no
 L^{\hi}_{\spp \ho}(\vec{p}^{\sp '}) &=& \frac{p^{\hi}}{m} + \frac{\d p^{\hi}}{m}, \quad
 L^{\ho}_{\spp \hi}(\vec{p}^{\sp '})  = -\frac{p_{\hi}}{m} - \frac{\d p_{\hi}}{m},
\eea
with $\d p^{\ha}$ given by \Eq{A5}.
To handle $L^i_{\sp j}(\vec{p}^{\sp '})$ we note that to $\O(\d \vec{p})$ (i.e $\O(d\tau)$) we have
$|\vec{p}^{\sp '}|^2 = |\vec{p} + \d \vec{p}| \approx |\vec{p}|^2 + 2 \vec{p}\cdot\d \vec{p}$
so that
$$
\frac{1}{|\vec{p}^{\sp '}|^2} \approx \frac{1}{|\vec{p}\,|^2} \left( 1-  \frac{2 \vec{p}\cdot\d \vec{p}}{|\vec{p}\,|^2}\right).
$$
Therefore
\bea{A9}
L^{\hi}_{\sp \hj}(\vec{p}^{\sp '}) &=&\delta^i_{\sp j} - \left(\frac{p^{\ho}}{m} + \frac{\d p^{\ho}}{m}-1\right)
\frac{(p^{\hi}+\d p^{\hi}) (p_{\hj}+\d p_{\hj})}{|\vec{p}\,|^2}
\left( 1-  \frac{2 \vec{p}\cdot\d \vec{p}}{|\vec{p}\,|^2}\right) \no
&=& \left(\delta^i_{\sp j} - (\gamma-1)\, \frac{p^i p_j}{|\vec{p}\,|^2} \right)
- (\gamma -1)\left(\frac{p^{\hi} \d p_{\hj} + \d p^{\hi} p_{\hj}}{|\vec{p}\,|^2}
- \frac{p^{\hi}p_{\hj}}{|\vec{p}\,|^2} \,\frac{2 \vec{p}\cdot\d\vec{p}}{|\vec{p}\,|^2}\right)
- \frac{p^{\hi}p_{\hj}}{|\vec{p}\,|^2}\,\frac{\d p^{\ho}}{m}.\hspace{1.5em}
\eea
Thus, separating out the first term in each of the equations \Eq{A8} and \Eq{A9} as composing $L(\vec{p})$,
and changing $\vec{p} \to -\vec{p}$ in \Eq{A7} to form $L^{-1}(\vec{p}) =L(-\vec{p})$
we have
\be{A10}
L^{-1}(\vec{p}^{\sp '}) \equiv L^{-1}(\vec{p}) + M(\vec{p})\,d\tau,
\ee
where we have defined the matrix $M(\vec{p})$ by
\bea{A11}
 M^{\ho}_{\spp \ho}(\vec{p}^{\sp '}) &=&  \frac{\db p^{\ho}}{m} \no
 M^{\hi}_{\spp \ho}(\vec{p}^{\sp '}) &=&  -\frac{\db p^{\hi}}{m}, \quad
 M^{\ho}_{\spp \hi}(\vec{p}) = \frac{\db p_{\hi}}{m}, \no
 M^{\hi}_{\sp \hj}(\vec{p}^{\sp '}) &=& -(\gamma -1)\left(\frac{p^{\hi} \db p_{\hj} + \db p^{\hi} p_{\hj}}{|\vec{p}\,|^2}
- \frac{p^{\hi}p_{\hj}}{|\vec{p}\,|^2} \,\frac{2 \vec{p}\cdot\db \vec{p}}{|\vec{p}\,|^2}\right)
- \frac{p^{\hi}p_{\hj}}{|\vec{p}\,|^2}\,\frac{\db p^{\ho}}{m}
\eea
and we have defined
\be{A12}
\db p^{\ha} \equiv \d p^{\ha}/d\tau = \lambda^{\ha}_{\spp \hb}\,p^{\hb}, \qquad \lambda_{\ha\hb} = - \lambda_{\hb\ha}.
\ee

We can now write out the Wigner rotation using \Eq{A4}, \Eq{A7} and \Eq{A10} as
\bea{A13}
W &=& L^{-1}(\vec{p}^{\sp '}) \, \Lambda \, L(\vec{p}) \no
&\approx& \big(L^{-1}(\vec{p}) + M(\vec{p})\,d\tau\big)\,\big( I + \lambda d\tau \big)\,L(\vec{p}) \no
&=& I + \big( L^{-1}(\vec{p})\,\lambda \, L(\vec{p}) + M(\vec{p})\, L(\vec{p}) \big)\,d\tau + \O(d\tau^2) \no
&\equiv& I + \vartheta \, d\tau + \O(d\tau^2),
\eea
where the infinitesimal Wigner rotation $\vartheta^{\ha}_{\spp \hb}$ is given by
\be{A14}
\vartheta^{\ha}_{\sp \hb} = \big[ L^{-1}(\vec{p})\,\lambda \, L(\vec{p}) + M(\vec{p})\, L(\vec{p}) \big]^{\ha}_{\spp \hb}\;,
\quad W^{\ha}_{\spp \hb} = \d^{\ha}_{\spp \hb} + \vartheta^{\ha}_{\spp \hb}\,d\tau.
\ee
We turn next to evaluating \Eq{A14} for the three cases (i) $\vartheta^{\ho}_{\spp \ho}$,
(ii) $\vartheta^{\hi}_{\sp \ho}$ and (iii) $\vartheta^{\hi}_{\sp \hj}$.

Consider the evaluation of  $\vartheta^{\ho}_{\spp \ho}$
\be{A15}
\vartheta^{\ho}_{\spp \ho} = L^{-1}(\vec{p})^{\ho}_{\spp \ha}\,\lambda^{\ha}_{\spp \hb} \, L(\vec{p})^{\hb}_{\spp \ho}
 + M(\vec{p})^{\ho}_{\spp \ha}\, L(\vec{p})^{\ha}_{\spp \ho},
 \ee
 which consists of two terms.
When expanded, the first term in \Eq{A15} becomes
\bea{A16}
L^{-1}(\vec{p})^{\ho}_{\spp \ha}\,\lambda^{\ha}_{\spp \hb} \, L(\vec{p})^{\hb}_{\spp \ho}
  &=& L^{-1}(\vec{p})^{\ho}_{\spp \ho}\,\lambda^{\ho}_{\spp \hi} \, L(\vec{p})^{\hi}_{\spp \ho}
  +  L^{-1}(\vec{p})^{\ho}_{\spp \hi}\,\lambda^{\hi}_{\spp \ho} \, L(\vec{p})^{\ho}_{\spp \ho}
  +  L^{-1}(\vec{p})^{\ho}_{\spp \hi}\,\lambda^{\hi}_{\spp \hj} \, L(\vec{p})^{\hj}_{\spp \ho}, \no
&=& \gamma\,\lambda^{\ho}_{\spp \hi}\,\frac{p^{\hi}}{m} + \frac{p_{\hi}}{m}\,\lambda^{\hi}_{\sp \ho}\,\gamma
+ \frac{p_{\hi}}{m}\,\lambda^{\hi}_{\sp \hj}\,\frac{p^{\hj}}{m}.
\eea
The last term in \Eq{16} $\lambda_{\hi\hj} \, p^{\hi} p^{\hj}/m^2$, vanishes due to the anti-symmetry of $\lambda_{\hi\hj}$.
The second term in \Eq{16} can put in the form of the negative of the first term with the manipulations
$$
p_{\hi}\, \lambda^{\hi}_{\sp \ho} = p^{\hi}\, \lambda_{\hi\ho} = - p^{\hi}\, \lambda_{\ho\hi}
= - \lambda^{\ho}_{\spp \hi}\,p^{\hi},
$$
and therefore, \Eq{A16} vanishes.

The second term of \Eq{A15} can be expanded to
\bea{A17}
M(\vec{p})^{\ho}_{\spp \ha}\, L(\vec{p})^{\ha}_{\spp \ho}
&=& M(\vec{p})^{\ho}_{\spp \ho}\, L(\vec{p})^{\ho}_{\spp \ho}
+ M(\vec{p})^{\ho}_{\spp \hi}\, L(\vec{p})^{\hi}_{\spp \ho}, \no
&=&
\frac{\db p^{\ho}}{m} \, \gamma + \frac{\db p_{\hi}}{m}\,\frac{p^{\hi}}{m}.
\eea
Using
\bea{A18}
\db p^{\ho} &=& \lambda^{\ho}_{\sp \ha}\,p^{\ha} = \lambda^{\ho}_{\sp \hi}\,p^{\hi}, \no
\db p_{\hi} &=& \lambda_{\hi}^{\sp \ha}\,p_{\ha} = \lambda_{\hi}^{\sp \ho}\,p_{\ho}
+ \lambda_{\hi}^{\sp \hj}\,p_{\hj},
\eea
\Eq{A17} takes the form
\be{A19}
\lambda^{\ho}_{\sp \hi}\,\frac{p^{\hi}}{m}\,\gamma + \frac{1}{m}\,[ \lambda_{\hi}^{\sp \ho}\,p_{\ho}
+ \lambda_{\hi}^{\sp \hj}\,p_{\hj} ]\,\frac{p^{\hi}}{m}.
\ee
Again the third term vanishes due to the anti-symmetry of $\lambda_{\hi\hj}$. Using $p_{\ho} = m\gamma$ and
$\lambda_{\hi}^{\sp \ho} = -\lambda^{\ho}_{\spp \hi}$, the second term cancels the first term, and \Eq{A19}
vanishes as well. The results of \Eq{A16} and \Eq{A17} then shows that $\vartheta^{\ho}_{\spp \ho} = 0$ as
stated in \Eq{A2}.

The evaluation of $\vartheta^{\hi}_{\sp \ho}$ and $\vartheta^{\hi}_{\sp \hj}$ proceed in an analogous fashion,
albeit with considerably more algebra which, though lengthy, is straightforward. The following relationships
prove useful in manipulating these expressions into the form given in \Eq{A2}. Note that
$\vec{p}\cdot\db\vec{p} \equiv \d_{\hi\hj}\,p^{\hi}\,\d p^{\hj} = -p_{\hi} \d p^{\hi}$, where in the last line
we have used the fact that $p^{\hi} = -p_{\hi}$ in the flat spacetime metric $\eta_{\hi\hj}$. Expanding the
normalization of $p^{\,' \ha}$, namely $(p^{\,' \ho})^2 = |\vec{p}^{\sp '}|^2 + m^2$ with
$p^{\,' \ha} = p^{\ha} + \db p^{\ha}$ one easily derives the useful relationship
$p^{\ho}\,\db p^{\ho} = \vec{p}\cdot\db\vec{p}$. Finally, the normalization of the unprimed momentum
$(p^{\ho})^2 = |\vec{p}\,|^2 + m^2$ can be put into the form of a useful identity  $m^2 (\gamma^2-1)/|\vec{p}\,|^2 = 1$
using $p^{\ho} = m \gamma$. These relations, along with the anti-symmetry of $\lambda_{\hi\hj}$ and
$p^{\ho} = p_{\ho}$ and $p^{\hi} = -p_{\hi}$ reveal after some algebraic effort that
$\vartheta^{\hi}_{\sp \ho} = \vartheta^{\ho}_{\spp \hi}=0$ and that $\vartheta^{\hi}_{\sp \hj}$ takes the
non-zero anti-symmetric form of a rotation matrix given in \Eq{A2} (\Eq{26} in the text).
\section{Derivation of the $\O(\lowercase{\hbar})$ corrected 4-acceleration \Eq{70}}\label{appB}
We wish to derive the quantum mechanical correction to the 4-acceleration $a^\a(x)$ in \Eq{70} from
the corresponding correction to the 4-velocity $v^\a(x)$ given in \Eq{67}. Since we are only
interested in keeping terms to $\O(\hbar)$ let us write
\be{C1}
v_\a(x) \equiv u_a(x) + \frac{\hbar}{2 m i} \d v_\a(x) + \O(\hbar^2), \quad a^\a(x) \equiv \hbar\d a^\a(x) + \O(\hbar^2),
\ee
where
\be{C2}
\d v_\a(x) = \bar{\psi}^{(\sigma)}_0(x) D_\a \psi^{(\sigma)}_0(x)
                                             - \big(D_\a \bar{\psi}^{(\sigma)}_0(x)\big) \, \psi^{(\sigma)}_0(x).
\ee
From \Eq{69} we have to $\O(\hbar)$
\bea{C3}
a_\a(x) &=& v^\b(x)\,D_\b v_\a(x) = 2 v^\b(x)\,D_{[\b} \,v_{\a]}(x), \no
&=& u^\b(x) 2 D_{[\b} \,u_{\a]}(x)
+  \frac{\hbar}{ m i}\,\left[\, \d v^\b(x) D_{[\b}\, u_{\a]}(x) + u^\b(x)\,D_{[\b}\,\d v_{\a]}(x) \,\right]
\eea
Consider the expression $2 D_{[\b} \,u_{\a]}(x)$ which appears in both the first and second terms in \Eq{C3}
\be{C4}
2 D_{[\b} \,u_{\a]}(x) = 2\nabla_{[\b} \,u_{\a]}(x) =  [\pd{\b}, u_{\a}(x)] = -[\pd{\b}, \pd{\a}S(x)] = 0.
\ee
In the first equality we have used the property that the action of the total covariant derivative $D_\b$  on a world vector is just the
Riemann covariant derivative $\nabla_\b$, while in the second we have used the symmetry of the Christoffel symbols
in their lower two covariant indices. Finally, we have used \Eq{53} which states that the 4-velocity
 is the normal to surfaces of constant action $u_\a(x) = -\pd{\a} S(x)$. Alternatively, note that the
 first term in \Eq{C3} is
 $$
u^\b(x) 2 D_{[\b} \,u_{\a]}(x) = u^\b(x) \nabla_{\b} \,u_{\a}(x) - u^\b(x)  \nabla_{\a} \,u_{\b}(x) = 0,
 $$
where the first term vanishes since $u^\a(x)$ is the tangent to the classical geodesic and the second
term vanishes since the normalization of the classical 4-velocity $u^\a(x)\,u_\a(x)=1$ implies that
$\nabla_\a \big(u^\b(x)\,u_\b(x)\big)=u^\b(x) \nabla_{\a} \,u_{\b}(x)=0$.

The remaining third term in \Eq{C3} can be expanded as
\bea{C5}
a_\a(x) &=& \frac{\hbar}{ m i}\, u^\b(x)\,D_{[\b}\,\d v_{\a]}(x), \no
&=& \frac{\hbar}{ m i}\, u^\b(x)
\left[\,
\bar{\psi}^{(\sigma)}_0(x) D_{[\b} D_{\a]} \psi^{(\sigma)}_0(x)
                                             - \big(D_{[\a} D_{\b]} \bar{\psi}^{(\sigma)}_0(x)\big) \, \psi^{(\sigma)}_0(x)
\,\right] \no
&+& \frac{\hbar}{ m i}\, u^\b(x)
\left[\,
D_{\b} \bar{\psi}^{(\sigma)}_0(x) \, D_{\a} \psi^{(\sigma)}_0(x)
                                             - \big( D_{\a} \bar{\psi}^{(\sigma)}_0(x)\big) \, D_{\b}\psi^{(\sigma)}_0(x)
\,\right].
\eea
The second line of the last equality in \Eq{C5} vanishes when we invoke the equation of motion
$u^\b(x) D_\b \psi^{(\sigma)}_0(x)$ which states that $\psi^{(\sigma)}_0(x)$ is parallel transported
along the $u^\b(x)$ congruence. In the first line of the second equality in \Eq{C5} we utilize the
expression for the commutator of the total covariant derivative acting on a spinor, which is
proportional to the Riemann curvature tensor \Eq{44} and its adjoint
\be{C6}
 [\, D_{\u}, D_{\v} \,] \psi(x) =  \spp \frac{i}{4} \, R_{\u\v\c\d}(\Gamma(x))\,\sigma^{\c\d}(x)\, \psi(x)
\ee
\be{C7}
[\, D_{\u}, D_{\v} \,] \bar{\psi}(x) =  -\frac{i}{4} \,R_{\u\v\c\d}(\Gamma(x))\,\bar{\psi}(x)\,\sigma^{\c\d}(x)
\ee
since $\sigma^{\c\d}(x)$ is a Hermetian matrix.
Substituting these last expressions into \Eq{C5} leads to our final expression for the 4-acceleration
\bea{C8}
a_\a(x) &=&  -\frac{\hbar}{4 m} \, R_{\a\b\hc\hd}(x) \,u^\b(x) \, \sigma^{\hc\hd}, \\
        &=&  -\frac{\hbar}{4 m} \, R_{\a\b\c\d}(x) \,u^\b(x) \, \sigma^{\c\d}(x) \label{C9}
\eea
using $\sigma^{\c\d}(x) = \e{c}{\c}(x)\,\e{d}{\d}(x)\,\sigma^{\hc\hd}$ and
$R_{\a\b\c\d}(x) = \inve{\c}{c}(x)\,\inve{\d}{d}(x)\,R_{\a\b\hc\hd}(x)$.

%
%




\begin{thebibliography}{99}
\bibitem{LLO_comment} Our figures were inspired and adapted from the wonderfully visusal local observer's laboratory (small box at
spacetime point $x$ defined by the 3-spatial 4-vector axes $\bfe{i}(x)$ and clock rate $\bfe{0}(x)$) in
Hartle's \tit{Gravity} (see Hartle in \cite{tetrads}). Here, we have redrawn the box in a simpler, less ornate
style, and used a cartoon image of one of the authors (PMA) as the local Lorentz observer measuring properties
of a particle passing through his local laboratory.
\end{thebibliography}

\end{document}